\documentclass[%
aps,reprint,
nofootinbib,amsmath,amssymb,amsfonts,
superscriptaddress
]{revtex4-1}

\usepackage{graphicx}
\usepackage{bm}
\usepackage{xcolor}
\usepackage[normalem]{ulem}

\usepackage[utf8]{inputenc}
\usepackage{mathalfa,amssymb}
\usepackage{tikz}
\usetikzlibrary{positioning}
\usetikzlibrary{shapes,arrows,arrows,positioning,fit}
\usepackage{mathrsfs}
\usepackage{comment}
\usepackage{caption}
\usepackage{subcaption}
\usepackage{amsmath} 
\usepackage{hyperref}
\usepackage{float}
\usepackage[section]{placeins}
\hypersetup{
    colorlinks=true,       
    linkcolor=red,          
    citecolor=blue,        
}

\begin{document}

\title{Analytical computation of quasi-normal modes of slowly-rotating black-holes in dCS gravity}
\author{Manu Srivastava}
\email{manu$_$s@iitb.ac.in}
\affiliation{Department of Physics, Indian Institute of Technology Bombay, Mumbai 400076, India} 
\author{Yanbei Chen} \email{yanbei@caltech.edu}
\affiliation{Theoretical Astrophysics 350-17, California Institute of Technology, Pasadena, California 91125}
\author{S. Shankaranarayanan} \email{shanki@phy.iitb.ac.in}
\affiliation{Department of Physics, Indian Institute of Technology Bombay, Mumbai 400076, India}

\date{\today}

\begin{abstract}
Using gravitational wave observations to search for deviations from general relativity in the strong-gravity regime has become an important research direction. 
One aspect of the strong gravity modifications to GR is parity violation. Chern Simons (CS) gravity is one of the most frequently studied parity-violating models of strong gravity. CS gravity is indistinguishable from GR for all conformally flat space-times and for space-times that possess a maximally symmetric 2-dimensional subspace. Also, it is known that the Kerr black-hole is not a solution for CS gravity. At the same time, the only rotating solution available in the literature for dynamical  CS (dCS) gravity is the slow-rotating case most accurately known to quadratic order in spin.
In this work, for the slow-rotating case (accurate to first order in spin), we derive the linear perturbation equations governing the metric and the dCS field accurate to linear order in spin and quadratic order in the CS coupling parameter ($\alpha$) and obtain the quasi-normal mode (QNM) frequencies. After confirming the recent results of Wagle et al. (2021), we find an \emph{additional contribution} to the eigenfrequency correction at the leading perturbative order of $ \alpha^2$. Unlike Wagle et al., we also find corrections to frequencies in the polar sector. We compute these extra corrections by evaluating the expectation values of the perturbative potential on unperturbed QNM wavefunctions along a contour deformed into the complex-$r$ plane. For $\alpha=0.1 M^2$, we obtain the ratio of the imaginary parts of the dCS correction to the GR correction in the first QNM frequency (in the polar sector) to be $0.263$ implying significant change. For the $(2,2)-$mode, the dCS corrections make imaginary part of the first QNM of the fundamental mode less negative, thereby decreasing the decay rate. Our results, along with future gravitational wave observations, can be used as a test for dCS gravity and to further constrain the CS coupling parameters.
\end{abstract}

\maketitle

\section{Introduction}
\label{sec:Intro}

When two compact objects (black-holes or neutron stars) coalesce to form a black-hole, the space-time geometry close to the remnant black-hole is highly distorted and radiates gravitational waves until it settles down to an equilibrium configuration~\cite{Regge:1957td,Zerilli,Vishveshwara:1970zz,Chandrasekhar:1985kt}. The gravitational wave signal has three distinct phases. During the {\it inspiral phase}, the two objects spiral in towards each other in a quasi-circular orbit, emitting GWs with increasing frequency with respect to time. The {\it merger phase} is when the two objects plunge towards each other and form a remnant compact object. In the {\it ring-down phase}, the remnant relaxes towards a stationary state and radiates all the perturbations away~\cite{Nollert:1999ji}. 
\vspace{\baselineskip}
\newline
Gravitational Waves (GWs) emitted during the ring-down phase are quasi-normal modes (QNMs) and offer valuable insight into the nature of the objects emitting them~\cite{Nollert:1999ji,Kokkotas:1999bd,Konoplya:2011qq}. The frequency and damping of these oscillations \emph{depend only} on the parameters characterizing the black-hole and are completely independent of the particular initial configuration that caused the excitation of such vibrations~\cite{Vishveshwara:1970zz}.  In General Relativity (GR), gravitational waves have two tensor polarization modes~\cite{1973-Misner.etal-Gravitation}. In generic metric theories of gravity, up to four additional polarizations can appear and may imply new effects~\cite{Barack:2018yly}. 
\vspace{\baselineskip}
\newline
The spectrum of QNM predicted by GR comprises of two isospectral towers of modes that are respectively even and odd under parity~\cite{Chandrasekhar:1985kt,Nollert:1999ji}. Besides extra polarization modes, alternative gravity theories can introduce three different effects on QNMs~\cite{Barack:2018yly}: First, to modify the spectrum of even and odd modes while preserving isospectrality. Second, to break isospectrality (do not emit GWs with equal energy in the two polarization states). Third, mix the even and odd modes, so that the eigenmodes do not have definite parity.
\vspace{\baselineskip}
\newline
Since the first discovery in 2015, a number of compact binary mergers have been detected from the data collected by Advanced LIGO and Advanced VIRGO~\cite{Abbott:2020niy,Bhagwat:2019bwv,Venumadhav:2019lyq}.   
These detections confirmed GR's predictions within their statistical uncertainty and enriched our understanding of the Universe by providing the first direct evidence of massive stellar-mass black holes and black holes colliding to form a single, larger black hole. \vspace{\baselineskip}
\newline
Although, the current LIGO-VIRGO detectors have only achieved signal-to-noise ratio around $5$ in the ring-down phase~\cite{LIGOSNR}, tests of GR involving QNMs have already been carried out. In particular, QNM overtones have been shown to offer initial tests of the ``no-hair'' theorem \cite{Isi:2019aib,Dhanpal:2018ufk,Cardoso_Request_Overtone_PhysRevD.97.044048}. 
%
%
The future gravitational-wave detectors (e.g., the Cosmic Explorer~\cite{Evans:2016mbw,Cardoso_Request_1_PhysRevLett.117.101102}) may detect GW signal-to-noise ratio in the quasi-normal mode regime ${\rm SNR} > 50$~\cite{Nakano:2015uja}, hence, will help to probe the QNM structure accurately and might provide more stringent test of GR in the strong gravity regime~\cite{Barack:2018yly, Berti_PhysRevD.101.024043, Berti_2_PhysRevD.100.044061}.  
\vspace{\baselineskip}
\newline
Naturally, there has been an intense theoretical activity to obtain unique gravitational wave signatures that distinguish GR and modified gravity theories. There are
many different ways to modify GR in the strong gravity regime, and each model has different observational signatures. A universal feature of any strong-gravity corrections to GR is introducing higher derivative Ricci scalar, Ricci tensor, and Riemann tensor terms in action~\cite{Stelle:1976gc,DeFelice:2010aj,Sotiriou:2008rp,Capozziello:2011et,deRham:2014zqa,Alexander:2009tp}. The main physical
motivations for these modifications of gravity consist of a possibility of a more realistic representation of the gravitational fields near curvature singularities and to create some first-order approximation for the models of quantum gravity~\cite{Donoghue:1994dn}.
\vspace{\baselineskip}
\newline
Will listed four basic criteria for the viability of a \emph{consistent gravity theory}~\cite{Will:2005va,Will:2018}: First, it must be complete. The theory should be able to analyze from {\sl first principles} the outcome of any experiment. Second, it must be self-consistent. Predictions should be unique and independent of the calculation method. Third, it must be relativistic: The theory should reduce to Special Relativity when gravity is {\sl turned off}. Fourth, it must have the correct Newtonian limit. In the weak gravitational fields and slow motion, it should reproduce Newton's laws.
\vspace{\baselineskip}
\newline
Chern-Simons (CS) modification to GR satisfies all the four criteria listed above~\cite{Jackiw:2003pm,Alexander:2009tp}. Chern-Simons Modified Gravity is a four-dimensional extension of GR that captures leading-order gravitational parity violation arising from the Pontryagin density~${}^{*}R\,R$. The Pontryagin
density is proportional to the wedge product $R \land R$ and a pseudo-scalar field $\vartheta$. CS theories are of two types: First is referred to as Canonical CS~\cite{Jackiw:2003pm}.
In this case, $ \vartheta $ is a constant and with no kinetic and potential term. The second is referred to as dynamical CS (dCS). 
In this case,  $ \vartheta $ is a fully dynamical field~\cite{Smith:2007jm}.
\vspace{\baselineskip}
\newline
For spherically symmetric space-times, the Pontryagin density vanishes, leading to standard GR with a scalar field potential. Owing to the no-hair theorem, for a spherically symmetric background, the only stable solution possible is that of the Schwarzschild. Hence, Schwarzschild space-time is a solution of both kinds of CS theories \cite{Jackiw:2003pm}. In the case of axisymmetric solutions, the Pontryagin density does not vanish, and hence, axisymmetric solutions are non-trivial to construct in CS theories~\cite{Delsate:2018ome}. So far in the literature, no closed form fast-spinning Kerr-like {analytic} solution exists in either kind of CS theories. However, it is possible to construct axisymmetric solutions from spherically symmetric solutions perturbatively in spin~\cite{Yunes:2009hc,Konno10.1143/PTP.122.561,Yagi_Spin2_PhysRevD.86.044037,Cano:2019ore}.  {Numerical solutions are obtained in Ref.~\cite{Delsate:2018ome}. More complex forms of CS coupling has also been considered, e.g., when the CS field $\vartheta$ couples quadratically to the Pontryagin density \cite{Gao:2018acg,Doneva:2021dcc}.}
\vspace{\baselineskip}
\newline
In the last few years, there is a lot of interest in studying the perturbations about Schwarzschild and
slowly-rotating black-holes in dCS gravity~\cite{Bhattacharyya:2018hsj,Okounkova:2019dfo,Okounkova:2019zjf,Wagle:2021tam,Cano:2020cao}. Interest has focused on two aspects:
First, is to confirm/infirm the isospectrality relation in dCS gravity~\cite{Barack:2018yly,Shankaranarayanan:2019yjx}. In Ref. \cite{Bhattacharyya:2018hsj}, the authors have shown that the isospectrality between odd and even parity perturbations is broken for a perturbed Schwarzschild black-hole and slowly rotating, in dCS gravity in a gauge-invariant manner. 
Second direction of interest is to obtain the QNM frequencies corresponding to the odd and even parity perturbations. In the recent work~\cite{Wagle:2021tam}, Wagle et al computed the QNM frequencies for the slowly rotating black-hole solution of dCS gravity derived in Ref.~\cite{Yunes:2009hc}. The authors derived the perturbation equations accurate to linear order in spin and linear-order coupling parameter $\alpha$ for even, odd, and the pseudo-scalar field $(\vartheta)$. They then employed numerical techniques to integrate the perturbation equations simultaneously to calculate the QNM frequencies. They found that the axial sector QNM frequencies were corrected at order $\alpha^2$ when compared to the QNM frequencies for a slowly rotating Kerr background, while there were no corrections to the polar sector QNM frequencies from the GR case. The authors claimed that to get the QNM frequencies accurate to $O(\alpha^2)$, one needs to consider perturbation equations accurate to $O(\alpha)$ only, and not to $O(\alpha^2)$. 
\vspace{\baselineskip}
\newline
%
In this work, we compute the QNM frequencies for the same slowly rotating solution of dCS gravity \cite{Yunes:2009hc}. There are \emph{few key differences} between the approach in this work and in Ref. \cite{Wagle:2021tam}. Here, we derive the perturbation equations accurate to quadratic order in $\alpha$ (and linear order in spin). Unlike what is claimed by the authors in \cite{Wagle:2021tam}, we find that the $O(\alpha^2)$ terms in the perturbation equations bring an additional correction to the QNM frequency correction at the leading perturbative order of $\alpha^2$. We get QNM frequency corrections even for the polar sector, unlike the results of \cite{Wagle:2021tam}. Another point of difference in our work is the procedure we use to calculate the QNM frequencies. We use an analytical technique to compute the QNM frequencies. Specifically, we use a procedure similar to that used in non-degenerate perturbation theory for Schroedinger equations.
We treat the terms independent of spin (corresponding to the Schwarzschild background) as the zeroth order terms in perturbation theory, while the terms proportional to spin (which include both GR and dCS  terms) are treated as a perturbing potential. The QNM frequency corrections are calculated by evaluating the expectation value of this perturbing potential between unperturbed QNM eigenfunctions (mode-functions for Schwarzschild background). We do so by analytically continuing the unperturbed eigenfunctions to the complex-$r$ plane and computing an integral along a suitably chosen contour deformed into the complex-$r$ plane. {For $\alpha=0.1 M^2$, we find the ratio of the imaginary parts of the dCS correction to the purely GR correction in the first QNM frequency (in the polar sector) to be $0.263$ implying significant change. Also, for $m\chi>0$, the dCS corrections make the magnitude of the imaginary part of the first QNM of the fundamental mode smaller, thereby decreasing the decay rate.}
\vspace{\baselineskip}
\newline
The paper is organised as follows: In section \ref{sec:CS_Intro}, we give some essential details of dCS gravity and introduce the slowly rotating black-hole solution that we will study in subsequent sections. In section \ref{sec:BH_Perturbation}, we perturb the black-hole background and compute the perturbation equations relevant to calculation of the QNM frequencies. In section \ref{sec:QNM_Calc}, we first do a Fermi estimate analysis to highlight that $O(\alpha^2)$ terms in perturbation equations do contribute to the $O(\alpha^2)$ corrections in QNM frequencies. We then go on to elaborate on our procedure of calculating the QNM frequency corrections, describing the perturbative scheme, defining the relevant inner product and choosing a contour for integration. We finally display our results, explicitly tabulating the corrections to QNM frequencies. Section \ref{sec:Discussion} contains a brief summary of the entire paper and discusses some directions of future work. The Appendices \ref{sec:Coeff_App}, \ref{sec:X_cos_W_cos}, \ref{sec:Inner_Product_Defn} and 
\ref{app:Fermi_Estimate} contain details of some of the discussions in Sec.~II to IV.
\vspace{\baselineskip}
\newline
We use $(-, +, +, +)$ signature for the 4-D space-time metric~\cite{1973-Misner.etal-Gravitation}. We use the geometric units $ G = c =1$ and $\kappa = 1/(16 \pi)$. We use the notations used in \cite{Kojima-PhysRevD.46.4289,Pani:2013pma,Wagle:2021tam} for convenient comparison. Our convention for the Levi Civita Tensor is $\epsilon^{1 2 3 4}=+1/{\sqrt{-g}}$. This leads to an overall sign difference in the Chern Simons field in our work compared to Ref. \cite{Yunes:2009hc}.
\section{Slow Rotating Black-hole Solution in dCS Gravity}
\label{sec:CS_Intro}
Chern Simons (CS) Modified Gravity is a parity-violating modification to Einstein's Gravity and was first postulated by Jackiw, and Pi \cite{Jackiw:2003pm}. The electromagnetic, strong, and gravitational interactions respect parity. So parity is a good symmetry for these interactions and is said to be conserved by them. However, Weak interaction does not respect parity. It is still unknown how parity violation arises from a unified field theory, including gravity. In principle, the parity violation in General Relativity leads to leptogenesis by transmitting itself into Baryon-Lepton violation through primordial gravity waves~\cite{Alexander:2009tp,Alexander:2004us}. Thus, CS gravity can potentially solve a few long-standing problems in particle physics and cosmology~\cite{Alexander:2009tp}. This section provides a quick review of dCS gravity and the slowly-rotating black-hole solution in this theory~\cite{Yunes:2009hc}. 
\subsection{dCS Gravity}
The dCS action is given by:
\begin{equation}
\label{eq:dCSaction}
S:=S_{\mathrm{EH}}+S_{\mathrm{CS}}+S_{\vartheta}+S_{\mathrm{mat}}\end{equation}
where $S_{EH}$ is the standard Einstein-Hilbert action:
\begin{eqnarray}
\label{eq:EHaction}
S_{\mathrm{EH}} &=& 
\kappa \int_{\mathcal{V}} d^{4} x \sqrt{-g} R \\
\label{eq:thetaaction}
S_{\vartheta} &=& \frac{-\beta}{2} \int_{\mathcal{V}} d^{4} x \sqrt{-g}\left[g^{a b}\left(\nabla_{a} \vartheta\right)\left(\nabla_{b} \vartheta\right)+2 V(\vartheta)\right] \\
\label{eq:couplingterm}
S_{\mathrm{CS}} &=& \frac{\alpha}{4} \int_{\mathcal{V}} d^{4} x \sqrt{-g} \vartheta^{*} R R
\end{eqnarray}
and $S_{mat}$ refers to the contribution from any other matter present in the spacetime:
%
Thus, dCS gravity contains an extra pseudo-scalar field $\vartheta$ whose action is identical to the canonical scalar field \eqref{eq:thetaaction}. The parity-violating term in action ($S_{CS}$) contains ${}^*RR$, which is called the Pontryagin density and is given by:
\begin{equation}{ }^{*} R R:={ }^{*}  R^{a}{ }_{b}{}^{c d} R^{b}{}_{a c d}\end{equation}
where the dual Riemann tensor is defined as
\begin{equation}{ }^{*} R^{a}{ }_{b}{}^{c d}:=\frac{1}{2} \epsilon^{c d e f} R^{a}{}_{b e f} \, ,
\end{equation}
with $\epsilon^{c d e f}$ being the Levi Civita tensor and $\vartheta$ being a function of space-time. The above dCS action reduces to GR for $\vartheta =~{\rm constant}$. 
%
%
Variation of the total action \eqref{eq:dCSaction} w.r.t the metric and scalar field, leads to the following equations of motion, respectively,
\begin{eqnarray}
\label{eqn:CS_EOM1}
G_{a b}+\frac{\alpha}{\kappa} C_{a b} &=& \frac{1}{2 \kappa} T_{a b} \\
\label{eqn:CS_EOM2}
\beta \square \vartheta &=& \beta \frac{d V}{d \vartheta}-\frac{\alpha}{4} \, {}^{*}R R
\end{eqnarray}
where
\begin{equation}T^{a b}=T^{ab}_{\rm mat}+T^{ab}_{\vartheta}=-\frac{2}{\sqrt{-g}}\left(\frac{\delta \mathcal{L}^{\mathrm{mat}}}{\delta g_{a b}}+\frac{\delta \mathcal{L}^{\vartheta}}{\delta g_{a b}}\right)\end{equation}
where  $\mathcal{L}$ are the respective Lagrangian densities and
\begin{equation}T_{a b}^{\vartheta}=\beta\left[\left(\nabla_{a} \vartheta\right)\left(\nabla_{b} \vartheta\right)-\frac{1}{2} g_{a b}\left(\nabla_{a} \vartheta\right)\left(\nabla^{a} \vartheta\right)-g_{a b} V(\vartheta)\right]\end{equation}
In this work, we assume all other matter fields are absent, and hence, set $T^{ab}_{\rm mat}=0$. We also set the pseudo-scalar field potential to zero ($V(\vartheta)=0$)~\cite{Yunes:2009hc}. The C-tensor in \eqref{eqn:CS_EOM1} is defined as:
\begin{equation}
\label{eqn:CTensor}
C^{a b}:=v_{c} \epsilon^{c d e(a} \nabla_{e} R^{b)}{}_{d}+v_{c d}\,{}^{*} R^{d(a b) c}\end{equation}
The braces in \eqref{eqn:CTensor} imply symmetrization of the indices. Also
\begin{equation}v_{a}:=\nabla_{a} \vartheta, \quad v_{a b}:=\nabla_{a} \nabla_{b} \vartheta=\nabla_{(a} \nabla_{b)} \vartheta\end{equation}
As mentioned earlier, the static spherically symmetric Schwarzschild solution of GR is also a solution of the CS theory because the Pontryagin density vanishes for the Schwarzschild geometry~\cite{Jackiw:2003pm}. On the other hand, the rotating Kerr solution of GR is not a solution of the CS theory. A rotating black hole solution with arbitrary angular momentum is unknown for the CS theory. However, Yunes and Pretorius proposed a slow rotating solution~\cite{Yunes:2009hc}. We will discuss the salient features of this solution in the following subsection.
\subsection{Slowly rotating black-hole solution}
\label{sec:slow_rot_CS_BH}

Since the dCS equations of motion \eqref{eqn:CS_EOM1}, \eqref{eqn:CS_EOM2} are highly coupled and non-linear, 
obtaining exact solutions are hard and one has to resort to approximations. In Ref.~\cite{Yunes:2009hc}, the authors solved equations \eqref{eqn:CS_EOM1} and \eqref{eqn:CS_EOM2} perturbatively to linear order in the spin parameter $a$ and quadratic order in the CS coupling parameter $\alpha$ and obtained the following slow-rotating solution:
\begin{equation}
\label{eqn:Slow-rotMetric}
d \bar{s}^{2} \simeq d s_{\mathrm{SR}}^{2}+\frac{5}{4} \frac{\alpha^{2}}{\beta\kappa} \frac{a}{r^{4}}\left(1+\frac{12}{7} \frac{M}{r}+\frac{27}{10} \frac{M^{2}}{r^{2}}\right) \sin ^{2} \theta d t d \varphi
\end{equation}
where $d s_{\mathrm{SR}}^{2}$ is the slow-rotating Kerr line element (up to linear order in $a$):
\begin{equation}
\label{eqn:slo-rotKerr}
d s_{\mathrm{SR}}^{2} = -f(r) d t^{2}-\frac{4 M a \sin ^{2} \theta}{r} d t d \varphi+\frac{d r^{2}}{f(r)}+r^{2} d\Omega^2
\end{equation}
where $f(r)=1-2M/r$. The background dCS field in the slow rotating solution is given by:
\begin{equation}
\label{eqn:Background_Theta}
\bar{\vartheta} \simeq -\frac{5}{8} \frac{a \alpha}{\beta M} \frac{\cos \theta}{r^{2}}\left(1+\frac{2 M}{r}+\frac{18 M^{2}}{5 r^{2}}\right)
\end{equation}
The above black-hole solution has two properties that will be useful for computing QNM frequencies: First, the horizon of the above solution is the same as that in GR.
Second, since the correction in metric is $r^{-4}$; at infinity, it is the same as slowly rotating Kerr. Hence, ADM mass and ADM angular momentum are the same as in GR. Note that the overall negative sign in Eq.~\eqref{eqn:Background_Theta} is different from \cite{Yunes:2009hc} due to the Levi Civita convention in this paper. As mentioned above, Ref. \cite{Yunes:2009hc} used a perturbation approach to obtain the above solution. As we will see below, the above expression for $\vartheta$ accurate to $O(\alpha)$ is sufficient to obtain the QNM frequencies accurate to $O(\alpha^2)$. In all the analytical calculations, the dCS parameter $\beta$ is arbitrary. However, in obtaining the numerical result in Sec. \ref{sec:QNM_Results}, we set $\beta = 1$.   
\vspace{\baselineskip}
\newline
In the next section, we set up the formal expressions to obtain the quasi-normal mode frequencies corresponding to this black-hole solution and \emph{highlight the differences} in our approach  Ref. \cite{Wagle:2021tam}.
\vspace{\baselineskip}

\section{Linear Perturbation about the slowly rotating black-hole solution}
\label{sec:BH_Perturbation}
In this section, we obtain the linear perturbations about the slowly rotating background black-hole space-time described in Section \ref{sec:slow_rot_CS_BH}. [For easy comparison, we use the notations used in Refs.~\cite{Kojima-PhysRevD.46.4289,Pani:2013pma,Wagle:2021tam}.]
The first order perturbations about the black-hole background is given by:
\begin{equation}
\label{eqn:MetricThetaPerturb}
g_{\mu \nu}=\bar{g}_{\mu \nu}+\epsilon \delta g_{\mu \nu}, \quad \vartheta=\bar{\vartheta}+\epsilon \delta \vartheta
\end{equation}
Based on their transformation under parity --- under the simultaneous transformations of $\theta \rightarrow \pi-\theta$ and $\varphi \rightarrow \pi + \varphi$ ---
the metric perturbations can be further divided into odd (axial) and even (polar) parity perturbations as:
\begin{equation}
    \delta g_{\mu \nu}(t, r, \theta, \varphi)=\delta g_{\mu \nu}^{\text {odd }}(t, r, \theta, \varphi)+\delta g_{\mu \nu}^{\text {even }}(t, r, \theta, \varphi)
\end{equation}
It is easy to see that under parity, the even and odd perturbations transform as $\delta g_{\mu \nu}^{\text {odd }} \rightarrow (-1)^{l+1}\delta g_{\mu \nu}^{\text {odd }}$ and $\delta g_{\mu \nu}^{\text {even }} \rightarrow (-1)^{l}\delta g_{\mu \nu}^{\text {even }}$. We work in the Regge-Wheeler gauge in which the metric perturbations can be written as (assuming harmonic time dependence):
\begin{equation}
\label{eqn:odd_perturb}
\delta g_{\mu \nu}^{\text {odd }}=\left(\begin{array}{cccc}
0 & 0 & h_{0}^{l m}(r) S_{\theta}^{l m}(\theta, \varphi) & h_{0}^{l m}(r) S_{\varphi}^{l m}(\theta, \varphi) \\
* & 0 & h_{1}^{l m}(r) S_{\theta}^{l m}(\theta, \varphi) & h_{1}^{l m}(r) S_{\varphi}^{l m}(\theta, \varphi) \\
* & * & 0 & 0 \\
* & * & * & 0
\end{array}\right)e^{-i \omega t}
\end{equation}
and
\begin{widetext}
\begin{equation}
\label{eqn:even_perturb}
\delta g_{\mu \nu}^{\text {even }}=\left(\begin{array}{cccc}
H_{0}^{l m}(r) Y_{l m}(\theta, \varphi) & H_{1}^{l m}(r) Y_{l m}(\theta, \varphi) & 0 & 0 \\
* & H_{2}^{l m}(r) Y_{l m}(\theta, \varphi) & 0 & 0 \\
* & * & r^{2} K^{l m}(r) Y_{l m}(\theta, \varphi) & 0 \\
* & * & * & r^{2} \sin ^{2} \theta K^{l m}(r) Y_{l m}(\theta, \varphi)
\end{array}\right)e^{-i \omega t}
\end{equation}
\end{widetext}
where $Y_{l m}(\theta, \varphi)$ are the spherical harmonics, 
\begin{equation}
\begin{split}
&S_{\theta}^{l m}(\theta, \varphi)=-\frac{1}{\sin \theta} \,  \partial_{\varphi} Y_{l m}(\theta, \varphi);\\&
S_{\varphi}^{l m}(\theta, \varphi)=\sin \theta \, \partial_{\theta} Y_{l m}(\theta, \varphi)
\end{split}
\end{equation}
and $*$ in the matrix elements denote symmetric components.
Note that in Eq.~\eqref{eqn:odd_perturb} and Eq.~\eqref{eqn:even_perturb}, there is an implicit summation over $m$ and $l$. $l \geqslant 0$ and $|m| \leqslant l$. The perturbation in the pseudo-field is decomposed into spherical harmonics as:
\begin{equation}
\delta \vartheta(t, r, \theta, \varphi)=\frac{R_{l m}(r)}{r} Y_{l m}(\theta, \varphi)e^{-i \omega t}
\end{equation}
\subsection{Angular decomposition in Regge-Wheeler gauge}
\label{sec:Kojima_Decomp}

The angular decomposition of the $10$ metric perturbation equations in the Regge-Wheeler gauge for a Schwarzschild background in GR was first shown by Kojima~ \cite{Kojima-PhysRevD.46.4289}. The analysis has been extended to slowly-rotating black-hole solutions in GR~\cite{Pani:2013pma}. In this subsection, we show a 
similar decomposition can be carried out for the 11 --- 10 metric $\delta g_{\mu\nu}$ and one scalar field $\vartheta$ --- perturbation equations for the slow rotating black-hole background up to $\alpha^2$. These 11 equations can be divided into three groups. 
\vspace{\baselineskip}
\newline
Group 1: From Eq.~\eqref{eqn:CS_EOM2} and the $tt$, $tr$, $rr$ and sum of $\theta\theta$ and $\varphi\varphi$ components of \eqref{eqn:CS_EOM1}, we get:
\begin{widetext}
\begin{equation}
\label{eqn:Eq_Group1}
\sum_{l, m}\left\{\left(A_{l m}^{(I)}+\tilde{A}_{l m}^{(I)} \cos \theta\right) Y_{l m}+B_{l m}^{(I)} \sin \theta \partial_{\theta} Y_{l m}+C_{l m}^{(I)} \partial_{\varphi} Y_{l m}\right\}=0 \quad(I=0 \text { to } 4)
\end{equation}
\end{widetext}
where $A$'s, $\tilde{A}$'s, $B$'s and $C$'s are the linear combinations of $H_0$, $H_1$, $H_2$, $K$, $h_0$, $h_1$, $R$ and depend only on $r$ and $\omega$. (Note that $t$ dependence goes away because of the harmonic time-dependence.)
\vspace{\baselineskip}
\newline
Group 2: From the $t \theta$, $r \theta$, $t \varphi$, $r \varphi$ components of \eqref{eqn:CS_EOM1}, we have:
\begin{widetext}
\begin{eqnarray}
\label{eqn:Eq_Group2_1}
\sum_{l, m}\left\{\left[\alpha_{l m}^{(J)}+\tilde{\alpha}_{l m}^{(J)} \cos \theta\right] \partial_{\theta} Y_{l m}-\left[\beta_{l m}^{(J)}+\tilde{\beta}_{l m}^{(J)} \cos \theta\right] \frac{\partial_{\varphi} Y_{l m}}{ \sin \theta} +\eta_{l m}^{(J)}\left(\sin \theta Y_{l m}\right)+\xi_{l m}^{(J)} X_{l m}+\chi_{l m}^{(J)}\left(\sin \theta W_{l m}\right)\right\}
=0 & & ~~ \\
\label{eqn:Eq_Group2_2}
\sum_{l, m}\left\{\left[\beta_{l m}^{(J)}+\tilde{\beta}_{l m}^{(J)} \cos \theta\right] \partial_{\theta} Y_{l m}+\left[\alpha_{l m}^{(J)}+\tilde{\alpha}_{l m}^{(J)} \cos \theta\right] \frac{\partial_{\varphi} Y_{l m}}{ \sin \theta}
+\zeta_{l m}^{(J)}\left(\sin \theta Y_{l m}\right)+\chi_{l m}^{(J)} X_{l m}-\xi_{l m}^{(J)}\left(\sin \theta W_{l m}\right)\right\}
= 0 & &~~
\end{eqnarray}

\begin{equation}
\label{eqn:X_W_Def}
\mbox{where} \qquad X_{l m}=2 \partial_{\varphi}\left(\partial_{\theta}-\cot \theta\right) Y_{l m}, \quad W_{l m}=\left(\partial_{\theta}^{2}-\cot \theta \partial_{\theta}-\frac{1}{\sin ^{2} \theta} \partial_{\varphi}^{2}\right) Y_{l m}
\end{equation}
\end{widetext}
In both Eqs.~\eqref{eqn:Eq_Group2_1}, and \eqref{eqn:Eq_Group2_2}, $J$ can be $0$ or $1$. Note that  $\alpha$'s, $\tilde{\alpha}$'s, $\beta$'s, $\tilde{\beta}$'s, $\eta$'s, $\chi$'s, $\zeta$'s and $\xi$'s are linear combinations of $H_0$, $H_1$, $H_2$, $K$, $h_0$, $h_1$, $R$ and depend only on $r$ and $\omega$. (Here again, $t$ dependence goes away because of the harmonic time-dependence.)
\vspace{\baselineskip}
\newline
Group 3: From the $\theta\varphi$ and the subtraction of $\theta\theta$ and $\varphi\varphi$ components of \eqref{eqn:CS_EOM1}, we get:
\begin{widetext}
\begin{eqnarray}
\label{eqn:Eq_Group3_1}
\sum_{l, m}\left\{f_{l m} \partial_{\theta} Y_{l m}+g_{l m}\left(\partial_{\varphi} Y_{l m} / \sin \theta\right)+s_{l m}\left(\frac{X_{l m}}{\sin ^{2} \theta}\right)+\tilde{s}_{l m}\left(\frac{X_{l m}\cos \theta}{\sin ^{2} \theta}\right)+t_{l m}\left(\frac{W_{l m}}{\sin \theta}\right)+\tilde{t}_{l m}\left(\frac{W_{l m}\cos \theta}{\sin \theta}\right)\right\}=0 
& & ~~~\\
\label{eqn:Eq_Group3_2}
\sum_{l, m}\left\{g_{l m} \partial_{\theta} Y_{l m}-f_{l m}\left(\partial_{\varphi} Y_{l m} / \sin \theta\right)-t_{l m}\left(\frac{X_{l m}}{\sin ^{2} \theta}\right)-\tilde{t}_{l m}\left(\frac{X_{l m}\cos \theta}{\sin ^{2} \theta}\right)+s_{l m}\left(\frac{W_{l m}}{\sin \theta}\right)+\tilde{s}_{l m}\left(\frac{W_{l m}\cos \theta}{\sin \theta}\right)\right\}=0 & & ~~~
\end{eqnarray}
\end{widetext}
%
Here again, $s$'s, $\tilde{s}$'s, $t$'s, $\tilde{t}$'s, $f$'s and $g$'s are some linear combinations of $H_0$, $H_1$, $H_2$, $K$, $h_0$, $h_1$, $R$ and depend only on $r$ and $\omega$. 
\vspace{\baselineskip}
\newline
Before we proceed with the rest of the calculations, we want to highlight the differences between GR and dCS gravity for the above angular decomposition. For dCS gravity, the structure of equations \eqref{eqn:Eq_Group1}, \eqref{eqn:Eq_Group2_1} and \eqref{eqn:Eq_Group2_2} are the same as in GR. The first difference is the appearance of the $\tilde{s}_{l m}$ and $\tilde{t}_{l m}$ terms in Eqs. \eqref{eqn:Eq_Group3_1} and \eqref{eqn:Eq_Group3_2}. These terms do not occur for a slowly rotating Kerr background. The second difference is that in dCS gravity, we have an extra pseudo-scalar field equation that is not present in GR. The structure of the scalar field equation is such that it can be grouped with the first set of metric equations \eqref{eqn:Eq_Group1}. 
\vspace{\baselineskip}
\newline
Note that the above angular decomposition leads to equations in terms of  \{$H_0$, $H_1$, $H_2$, $K$, $h_0$, $h_1$, $R$\} and the angular functions summed over all values of $l$ and $m$. To eliminate the angular dependencies, we need to use the orthogonality properties of spherical harmonics to get equations purely in terms of the radial functions \{$H_0$, $H_1$, $H_2$, $K$, $h_0$, $h_1$, $R$\} for individual ($l\,m$)-modes. For instance, 
we can multiply the perturbation equations \eqref{eqn:Eq_Group1} by $Y^{*}_{l m}$ and integrate over the solid angle to eliminate the angular functions. Detailed procedure to eliminate the angular dependencies for GR case involves a series of steps which is discussed in Ref. \cite{Pani:2013pma}. 
Appendix \ref{sec:X_cos_W_cos} 
contains the detailed procedure to account for the extra terms $\tilde{s}_{l m}$ and $\tilde{t}_{l m}$ terms in Eqs. \eqref{eqn:Eq_Group3_1} and \eqref{eqn:Eq_Group3_2} for the dCS gravity case. Following the elimination procedures in \cite{Pani:2013pma} and Appendix \ref{sec:X_cos_W_cos} leads to 11 equations of the form: 
%
\begin{equation}
\label{eqn:General_Perturb_Eq_Mode_Coupling}
M_{l m}+m a \tilde{M}_{l m}+a\left(q_{l} \check{M}_{l-1, m}+q_{l+1} \check{M}_{l+1, m}\right)=0
\end{equation}
where the $M$'s, $\tilde{M}$'s and $\check{M}$'s are some linear combinations of $H_0$, $H_1$, $H_2$, $K$, $h_0$, $h_1$, $R$ (with the same $l\,m$ indices) and depend only on $r$ and $\omega$. Note that the prefactors of $\alpha^i$ ($i=0,1,2$) have been suppressed in Eq.~\eqref{eqn:General_Perturb_Eq_Mode_Coupling}.

\subsection{Decoupled perturbation equations}
\label{sec:Simplifying_Pert_Eq}

To obtain the QNM frequencies, we need to further simplify Eqs.~ \eqref{eqn:General_Perturb_Eq_Mode_Coupling}. This simplification is possible by looking at the symmetry properties of the Axial and Polar perturbation variables in Eqs. \eqref{eqn:General_Perturb_Eq_Mode_Coupling}. In Ref.~\cite{Wagle:2021tam}, the authors argued that
under simultaneous transformations
\begin{equation}
\label{eqn:simultaneous_transformation}
\begin{aligned}
x_{l, m} & \rightarrow \mp x_{l,-m}, & y_{l, m} & \rightarrow \pm y_{l,-m}, \\
m & \rightarrow-m, & a & \rightarrow-a,
\end{aligned}
\end{equation}
equations \eqref{eqn:General_Perturb_Eq_Mode_Coupling} remain invariant. Note that $x_{l m}$ and $y_{l m}$ represent the axial and polar perturbation variables respectively, with indices $(l, m)$. Under the above transformations, the boundary conditions for QNMs of slowly rotating $\mathrm{BHs}$ in $\mathrm{dCS}$ are also invariant. The QNM frequencies in the slow rotation limit must also remain invariant and hence should be of the form:
\begin{equation}
\label{eqn:omega_variation_form}
\omega=\omega_{0}+m a \omega_{1}+a \omega_{2}+\mathcal{O}\left(a^{2}\right)~~{\rm with}~~\omega_2 = 0
\end{equation}
This argument holds irrespective of the order of $\alpha$. 

\begin{figure*}[!htb]
    \centering
 	\tikzstyle{decision} = [diamond, draw, fill=blue!20, 
 	text width=5.5em, text badly centered, inner sep=0pt]
 	\tikzstyle{block} = [rectangle, draw, fill=blue!20, 
 	text width=17em, text centered, rounded corners, minimum height=2em]
 	\tikzstyle{line} = [draw, -latex']
 	\tikzstyle{cloud} = [draw, ellipse,fill=red!20,
 	minimum height=4em]
 	\tikzstyle{decisionn} = [diamond, draw, fill=red!20, 
 	text width=4em, text badly centered, node distance=4cm, inner sep=0pt]
 	\tikzstyle{blockk} = [rectangle, draw, fill=blue!20, 
 	text width=8em, text centered, rounded corners, minimum height=2em]
 	\tikzstyle{blockkk} = [rectangle, draw, fill=red!20, 
 	text width=22em, text centered, rounded corners, minimum height=2em]
 	
 	\begin{tikzpicture}[node distance = 2.1cm, auto]\label{ams1}
 	\node [block] (B11) {Step 1: To get the dCS field equation};
  	\node [block, below of=B11] (B12) {Consider the axial equation set \eqref{eqn:General_Axial_Perturb_Eq} till $O(\alpha^2)$ };
 	\node [block, below of=B12] (B13) {We solve for $h_0$ in terms of $h_1$, $R$ and their derivatives.};
 	\node [block, below of=B13] (B14) {We substitute the obtained $h_0$ in Eq \eqref{eqn:General_Theta_Perturb_Eq} to get an equation of $O(\alpha^2)$ in terms of $h_1$, $R$ and their derivatives.};
 	\node [block, below of=B14] (B15) {We expand the obtained equation to linear order in $\chi$ and quadratic order in $\alpha$. Redefine $h_1$ in terms of the Regge-Wheeler function \eqref{eqn:RW_Func}};
 	\node [block, below of=B15] (B16) {We obtain the Schroedinger-like dCS field equation \eqref{eqn:Theta_Eq}};
 	\node [blockkk, right of=B11, node distance=8.5cm] (B21) {Step 2: To get the axial and polar sector equations};
 	\node [blockkk, below of=B21] (B22) {Consider the axial set of 3 equations \eqref{eqn:General_Axial_Perturb_Eq} and the polar set of 7 equations \eqref{eqn:General_Polar_Perturb_Eq}. They contain higher order derivative terms at order $\alpha^2$. These have to be approximated by terms at one perturbative order lower.};
 	\node [blockkk, below of=B22] (B23) {We solve equation set \eqref{eqn:General_Axial_Perturb_Eq} and \eqref{eqn:General_Polar_Perturb_Eq} till $O(\alpha)$ to get the required higher order derivative terms in terms of lower order derivative terms.};
 	\node [blockkk, below of=B23] (B24) {We substitute these expressions for the higher order derivative terms back in the original equation sets \eqref{eqn:General_Axial_Perturb_Eq} and \eqref{eqn:General_Polar_Perturb_Eq} to get axial and polar equation sets till $O(\alpha^2)$ with no higher order derivative terms.};
 	\node [blockkk, below of=B24, node distance=3cm] (B25) {We employ the standard GR procedure, substituting for $\partial_{rr}R$ at each step from \eqref{eqn:Theta_Eq}};
 	\node [blockkk, below of=B25, node distance=2.5cm] (B26) {We obtain the axial \eqref{eqn:Axial_Eq} and the polar \eqref{eqn:Polar_Eq} sector Schroedinger-like equations.};
 	\path [line] (B11) -- (B12);
 	\path [line] (B12) -- (B13);
 	\path [line] (B13) -- (B14);
 	\path [line] (B14) -- (B15);
 	\path [line] (B15) -- (B16);
 	\path [line] (B21) -- (B22);
 	\path [line] (B22) -- (B23);
 	\path [line] (B23) -- (B24);
 	\path [line] (B24) -- (B25);
 	\path [line] (B25) -- (B26);
 	\path [line] (B16) -- (B25);
 	\end{tikzpicture}
 	\caption{Flow chart provides the detailed procedure to obtain Schroedinger-like equations for the slowly rotating background in dCS gravity} 	\label{fig:flowchart}
 \end{figure*}
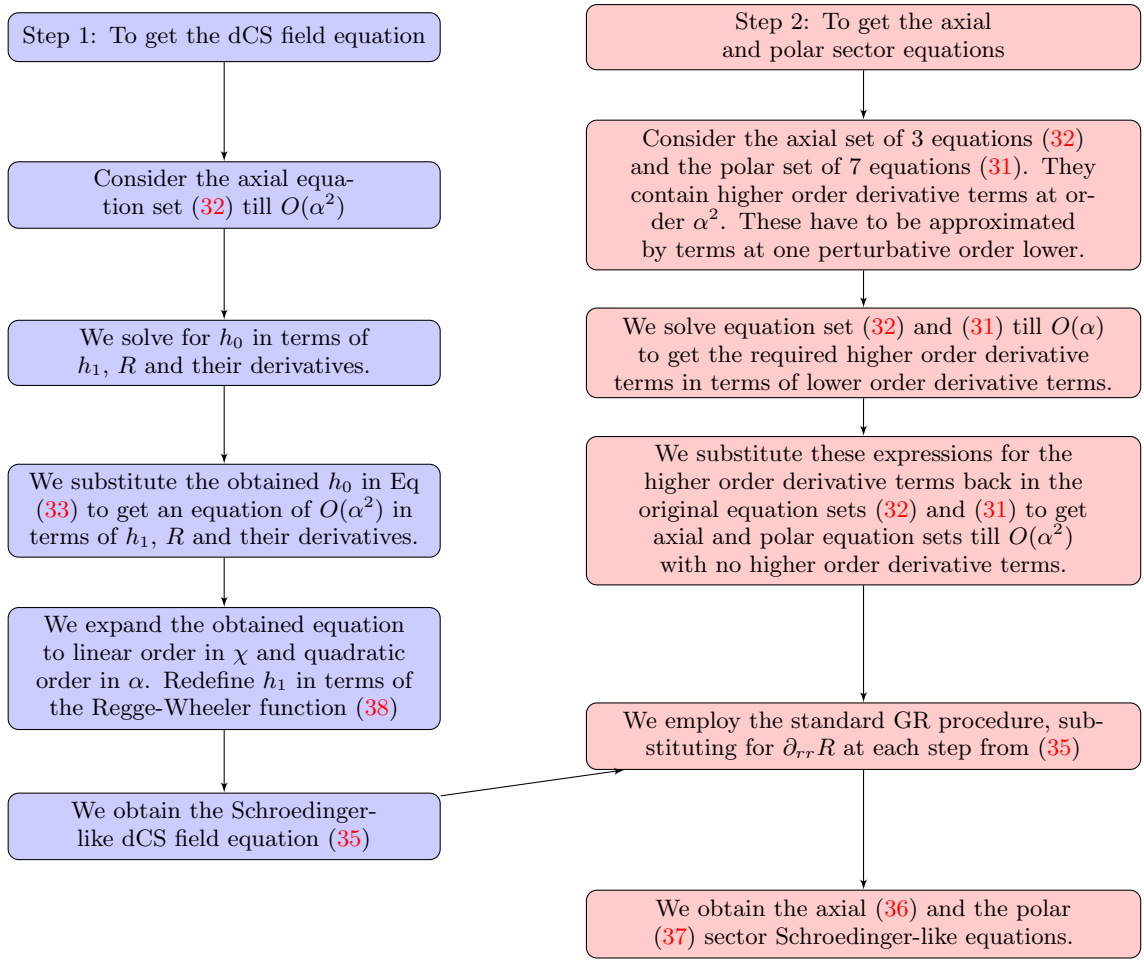

The immediate consequence of the above symmetry argument is that the quasi-normal mode (QNM) frequencies are not affected by mode couplings (of $l$ to $l\pm 1$) at leading order in spin (see Ref.~\cite{Wagle:2021tam} for details). 
%
The crux of the argument is that the coupling terms in \eqref{eqn:General_Perturb_Eq_Mode_Coupling} can only contribute to $\omega_2$ part of the frequency correction in \eqref{eqn:omega_variation_form}. And since $\omega_2$ should be zero because equations in \eqref{eqn:General_Perturb_Eq_Mode_Coupling} and its boundary conditions don't change under the simultaneous transformations \eqref{eqn:simultaneous_transformation}, there is no effect of the coupling terms on QNM frequencies. The coupling terms cannot source $\omega_1$ correction because the coupling terms in \eqref{eqn:General_Perturb_Eq_Mode_Coupling} are just the Schwarzschild functions and therefore are independent of $m$, considering terms only to linear order in spin in Eq. \eqref{eqn:General_Perturb_Eq_Mode_Coupling}.
Hence, we neglect the mode coupling terms of $\check{M}$ in Eq.~\eqref{eqn:General_Perturb_Eq_Mode_Coupling} in the rest of our analysis. If we don't have the $\check{M}$ terms, and if we explicitly write the $\alpha$ dependence of the terms, then \eqref{eqn:General_Perturb_Eq_Mode_Coupling} leads to 7 equations of the form:
\begin{equation}
\label{eqn:General_Polar_Perturb_Eq}
P^{(I)}_{l m}+\alpha^{2}\hat{P}^{(I)}_{l m}=0; \quad(I=1 \text{ to } 7)
\end{equation}
Three equations of the form:
\begin{equation}
\label{eqn:General_Axial_Perturb_Eq}
A^{(J)}_{l m}+\alpha \tilde{S}^{(J)}_{l m}+ \alpha^{2}\hat{A}^{(J)}_{l m}=0; \quad(J=1 \text{ to } 3)
\end{equation}
and one equation of the form:
\begin{equation}
\label{eqn:General_Theta_Perturb_Eq}
S_{l m}+\alpha \tilde{A}_{l m}+ \alpha^{2}\hat{S}_{l m}=0
\end{equation}
where the $P$'s are linear combinations of the polar functions $H_0$, $H_1$, $H_2$, $K$ and their derivatives. The $A$'s are linear combinations of the axial functions $h_0$, $h_1$ and their derivatives, and the $S$'s are linear combinations of the scalar field function $R$ and its derivatives. In the above equations, we have suppressed the spin ($a$) dependence of terms. 
From the above expressions, we find that the polar sector \eqref{eqn:General_Polar_Perturb_Eq} is independent of the pseudo-scalar field, whereas the axial sector \eqref{eqn:General_Axial_Perturb_Eq} is coupled to the pseudo-scalar field \eqref{eqn:General_Theta_Perturb_Eq}. This is also a feature for spherically symmetric space-times~\cite{Bhattacharyya:2018hsj}.
\vspace{\baselineskip}
\newline
While the above equations \eqref{eqn:General_Polar_Perturb_Eq}, \eqref{eqn:General_Axial_Perturb_Eq} and \eqref{eqn:General_Theta_Perturb_Eq} do not have mode-coupling terms, they are not in a form that can be related to the two polarization modes of the gravitational waves and scalar field. 
In other words, we need to combine the seven equations in \eqref{eqn:General_Polar_Perturb_Eq}
to a single differential equation that is analogous to Zerilli equation \cite{Zerilli} in the case of Schwarzschild space-time. 
Similarly, we need to combine the three equations in \eqref{eqn:General_Axial_Perturb_Eq}
to a single differential equation that is analogous to Regge-Wheeler equation \cite{Regge:1957td} in the case of Schwarzschild space-time. However, unlike Schwarzschild background in GR, the axial modes here are coupled to the scalar field. Since the Regge-Wheeler and Zerilli equations are of Schroedinger form, we will refer to the final Polar, Axial and Scalar equations for the slowly-rotating case as  Schroedinger-like equations~\cite{Chandrasekhar:1985kt}. \vspace{\baselineskip}
\newline
The procedure to get the Schroedinger-like equations starting from \eqref{eqn:General_Polar_Perturb_Eq}, \eqref{eqn:General_Axial_Perturb_Eq} and \eqref{eqn:General_Theta_Perturb_Eq} is a tedious one involving several steps. The flowchart (Fig. \ref{fig:flowchart}) describes the entire procedure. In the rest of this subsection, we give a bird's eye view of the procedure without going into the details. 
%
\vspace{\baselineskip}
\newline
We start with the simplest of the three ---  Schroedinger-like equation for the pseudo-scalar field. Eq. \eqref{eqn:General_Theta_Perturb_Eq} is the equation for the scalar field which also contains the axial perturbations variables that are governed by equations \eqref{eqn:General_Axial_Perturb_Eq}. Note that, to get an equation accurate to $O(\alpha^2)$ in Eq. \eqref{eqn:General_Theta_Perturb_Eq}, we can neglect the $O(\alpha^2)$ terms in \eqref{eqn:General_Axial_Perturb_Eq}. This is because any term at $O(\alpha^2)$ in Eq.~\eqref{eqn:General_Axial_Perturb_Eq}, when substituted back in Eq.~\eqref{eqn:General_Theta_Perturb_Eq}, will contribute at $O(\alpha^3)$. Therefore we solve the system of Eqs.~\eqref{eqn:General_Axial_Perturb_Eq} for some relations between $h_0$, $h_1$ and $R$ accurate to linear order in $\alpha$ and substitute them in \eqref{eqn:General_Theta_Perturb_Eq}. This leads to Schroedinger-like equation for the pseudo-scalar field perturbation $R_{l m}$ (cf. Eq.~\eqref{eqn:Theta_Eq} in Sec.~\ref{sec:Schroedinger-like_Eq}).
\vspace{\baselineskip}
\newline
We now focus on the polar \eqref{eqn:General_Polar_Perturb_Eq} and axial sectors \eqref{eqn:General_Axial_Perturb_Eq}. The many-step procedure of computing the Schroedinger-like equations for the polar and the axial sectors for the slowly rotating Kerr case is discussed in Ref.~\cite{Pani:2013pma}. In this work, we refer to this procedure as \emph{standard GR procedure}.  While Wagle et al. ~\cite{Wagle:2021tam} considered perturbation equations accurate to first-order in $\alpha$, in this work, we derive the equations accurate to $O(\alpha^2)$. $O(\alpha^2)$
correction terms in the perturbation equations, 
have higher-order derivative terms that are absent in the equations accurate to first-order in $\alpha$. Hence, we cannot directly use the standard GR procedure in this case. The modification to the procedure is explained as Step 2 in the flowchart (Fig. \ref{fig:flowchart}).
\vspace{\baselineskip}

\subsection{Schroedinger-like equations for the pseudo-scalar field, axial and polar sectors}
\label{sec:Schroedinger-like_Eq}
Following the procedure discussed in the flowchart (Fig. \ref{fig:flowchart}), we obtain the Schroedinger-like equations for the pseudo-scalar field perturbation, the axial sector, and the polar sector. We rewrite the equations in dimensionless variables $\chi, \tilde{\alpha}$:
\begin{equation}
    \chi=\frac{a}{M}\quad\quad;\quad\quad\quad \Tilde{\alpha}=\frac{\alpha}{M^2}
\end{equation}
The linear order perturbation equation for the pseudo-scalar field, accurate up to $O\left(\Tilde{\alpha}^2\right)$, is:
\begin{widetext}
\begin{equation}
\label{eqn:Theta_Eq}
\begin{array}{rrr}
f(r)^{2} \partial_{r r} R_{l m}+\frac{2 M}{r^{2}} f(r) \partial_{r} R_{l m}+\Tilde{\alpha}^2\,\chi\, m\, T_{1}(r) \partial_{r} R_{l m}+\left[\omega^{2}-V_{\mathrm{eff}}^{Smod}(r, \chi, \Tilde{\alpha}^{2})\right] R_{l m}
&\\
=-\Tilde{\alpha}\, f(r)\left\{[g(r)+\chi m\, h(r)] \Psi_{l m}^{\mathrm{RW}}+\chi m\, j(r) \partial_{r} \Psi_{l m}^{\mathrm{RW}}\right\}
\end{array}
\end{equation}
The linear order perturbation equation, accurate upto $O\left(\Tilde{\alpha}^2\right)$, for the Axial sector is:
\begin{equation}
\label{eqn:Axial_Eq}
\begin{array}{rrr}
f(r)^{2} \partial_{r r} \Psi_{l m}^{\mathrm{RW}}+\frac{2 M}{r^{2}} f(r) \partial_{r} \Psi_{l m}^{\mathrm{RW}}+\Tilde{\alpha}^2\,\chi\, m\, A_{1}(r) \partial_{r} \Psi_{l m}^{\mathrm{RW}}+\left[\omega^{2}-V_{\mathrm{eff}}^{Amod}(r, \chi, \Tilde{\alpha}^2)\right] \Psi_{l m}^{\mathrm{RW}} 
&\\
=-\Tilde{\alpha}\, f(r)\left\{[v(r)+\chi m\, n(r)] R_{l m}+\chi m\, p(r) \partial_{r} R_{l m}\right\}
\end{array}
\end{equation}
The linear order perturbation equation, accurate upto $O\left(\Tilde{\alpha}^2\right)$, for the
Polar sector is:
\begin{equation}
\label{eqn:Polar_Eq}
f(r)^{2} \partial_{r r} \Psi_{l m}^{\mathrm{Z}}+\frac{2 M}{r^{2}} f(r) \partial_{r} \Psi_{l m}^{\mathrm{Z}}+\Tilde{\alpha}^2\,\chi\, m\, P_{1}(r) \partial_{r} \Psi_{l m}^{\mathrm{Z}}+\left[\omega^{2}-V_{\mathrm{eff}}^{Pmod}(r, \chi, \Tilde{\alpha}^2)\right] \Psi_{l m}^{\mathrm{Z}}=0
\end{equation}
%
where the Regge-Wheeler function ($\Psi_{l m}^{\mathrm{RW}}$) describing axial perturbations and Zerilli-Moncrief function ($\Psi_{l m}^{\mathrm{Z}}$) describing the polar perturbation are
%
\begin{equation}
\label{eqn:RW_Func}
\Psi_{l m}^{\mathrm{RW}}=\frac{f(r)}{r}\left(1+\frac{2 m M \chi}{r^{3} \omega}\right) h_{1}^{l m}; \qquad
   \Psi_{l m}^{\mathrm{Z}}=\frac{2 r^2 \omega  \mathrm{K}^{l m}(r)+2 i(2 M-r) \mathrm{H}_1^{l m}(r)}{\omega  \left(\left(l^2+l-2\right) r+6 M\right)}(1+\chi\, \mathcal{C}(r))
\end{equation}
with
\begin{equation}
 \mathcal{C}(r)=\frac{2 m M^2 \left(-2 \left(l^2+l-2\right)^2 M r^2+\left(l^2+l-2\right) r^3 \left(l^2+l+2 r^2 \omega ^2-2\right)-24 M^2 r+48 M^3+12 M r^4 \omega ^2\right)}{l (l+1) r^4 \omega  \left(\left(l^2+l-2\right) r+6 M\right)^2} 
\end{equation}
\end{widetext}
and the potentials are:
\begin{eqnarray}
\label{eqn:Potential_Theta}
 V_{\mathrm{eff}}^{Smod}(r, \chi, \alpha^2) &=& V_{\mathrm{eff}}^{S}(r, \chi)+\Tilde{\alpha}^2\, T_2(r,\chi) \\
 \label{eqn:Potential_Axial}
    V_{\mathrm{eff}}^{Amod}(r, \chi, \alpha^2) & =& V_{\mathrm{eff}}^{A}(r, \chi)+\Tilde{\alpha}^2\,\chi\,m\,A_2(r) \\
    \label{eqn:Potential_Polar}
    V_{\mathrm{eff}}^{Pmod}(r, \chi, \alpha^2) &=& V_{\mathrm{eff}}^{P}(r, \chi)+\Tilde{\alpha}^2\,\chi\,m\,P_2(r)
\end{eqnarray}
The exact form of the above potential functions and all the coefficient functions that occur in \eqref{eqn:Theta_Eq}, \eqref{eqn:Axial_Eq} and \eqref{eqn:Polar_Eq} are given in Appendix \ref{sec:Coeff_App}. Equations (\ref{eqn:Theta_Eq} - \ref{eqn:Polar_Eq})
and (\ref{eqn:Potential_Theta} - \ref{eqn:Potential_Polar}) are accurate to quadratic order in $\alpha$ and contain $O(\alpha^2)$ terms that were not included in Ref.~\cite{Wagle:2021tam}. The Schroedinger-like form of the equations becomes apparent only if we express them in terms of the tortoise coordinate~($r_*$): $r_*=r+2M\log(r/2M\,-1)$.
\section{Evaluating the QNM frequencies}
\label{sec:QNM_Calc}
In this section, we evaluate the QNM frequencies for the three sectors analytically. Specifically, we use a procedure similar to that used in non-degenerate perturbation theory in Quantum Mechanics. It is important to highlight that only the technique is similar while the calculation is purely classical. We treat the terms independent of spin (corresponding to the Schwarzschild background) as the zeroth-order terms in perturbation theory. In contrast, the terms proportional to spin (including GR and dCS terms) are treated as a perturbing potential. 

Wagle et al. \cite{Wagle:2021tam} performed a Fermi estimation analysis to determine the dependence of the QNM frequency on the CS coupling parameter. To compare and contrast the results and highlight that some relevant perturbative terms are not present in Ref.~\cite{Wagle:2021tam}, we repeat the Fermi estimate analysis for the equations derived in section \ref{sec:Schroedinger-like_Eq} in Appendix \ref{app:Fermi_Estimate}. 

From Appendix \ref{app:Fermi_Estimate} we see that the results are different from those in Ref.~\cite{Wagle:2021tam}. Our analysis reveals an additional contribution ($(A_2-G_A)$ term) to the frequency correction compared to that in Ref.~\cite{Wagle:2021tam}. Thus, even using Fermi estimate, it is clear that the analysis in Ref. \cite{Wagle:2021tam} has not included certain terms that contribute to the correction in $\omega$ at $O(\Tilde{\alpha}^2)$. 
\subsection{Perturbation Scheme, Inner Product, and Contour}
\begin{figure}[!htb]
    \centering
    \vspace*{15pt}
    \includegraphics[scale=0.55]{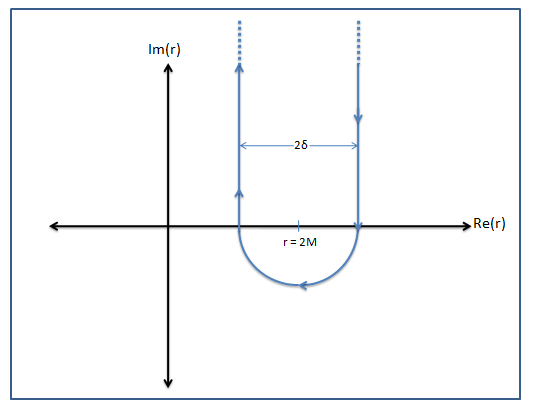}
    \caption{Contour for the Integral in Inner Product Definition}
    \label{fig:Contour}
\end{figure}

In the rest of this section, we discuss the analytical procedure we use to compute the corrections to the QNM frequencies. We employ a perturbation scheme in the spin parameter $\chi$ with the Schwarzschild case being the zeroth-order solution. We compute the Schwarzschild QNM eigenmodes and frequencies using the well-known continued fractions analysis by Leaver \cite{Leaver:1985ax}. 
\vspace{\baselineskip}
\newline
Since the Polar sector is uncoupled, we first discuss the polar sector \eqref{eqn:Polar_Eq}. Setting $\rho=-i\,\omega$,
Eq.~\eqref{eqn:Polar_Eq} can be rewritten as:
\begin{equation}
\label{eqn:Polar_Eq_Perturb_Theory}
    \hat{O}^P\Psi+\chi\,\hat{V}^P\Psi=\rho^2\Psi
\end{equation}
where
\begin{eqnarray}
\label{eqn:Polar_Operator}
\hat{O}^P &=& \frac{d^2}{dr_*^2}-V^P_{Sch} \\
\label{eqn:Polar_Perturb_Operator}
\hat{V}^P &=& -(V_{\mathrm{eff}}^{Pmod}-V^P_{Sch})+\tilde{\alpha}^2 m\, P_1(r) \frac{d}{dr}
\end{eqnarray}
and $V^P_{Sch}$ is the Zerilli potential which is obtained by setting $\chi=0$ in $V_{\mathrm{eff}}^{Pmod}$. 
\vspace{\baselineskip}
\newline
As mentioned above, our aim is to solve \eqref{eqn:Polar_Eq_Perturb_Theory} perturbatively and $\chi$ keeps track of the higher and higher order corrections to the Schwarzschild case. Following the standard non-degenerate perturbation theory, we expand $\Psi$ and $\rho$ in powers of $\chi$:
\begin{equation}
\begin{split}
    &\Psi=\Psi^{(0)}+\chi\Psi^{(1)}+\chi^2\Psi^{(2)}+...\\&
    \rho=\rho^{(0)}+\chi\rho^{(1)}+\chi^2\rho^{(2)}+... \, .
\end{split}
\end{equation}
We now substitute these expansions back in \eqref{eqn:Polar_Eq_Perturb_Theory}, expand and 
collect terms at each order of $\chi$, and set coefficients 
at each order of $\chi$ to zero. We get at zeroth order:
\begin{equation}
    \hat{O}^P\Psi^{(0)}=(\rho^{(0)})^2\Psi^{(0)}
\end{equation}
and at the first order:
\begin{equation}
\label{eqn:first_order_in_Chi_Polar}
    \hat{O}^P\Psi^{(1)}+\hat{V}^P\Psi^{(0)}=2\rho^{(0)}\rho^{(1)}\Psi^{(0)}+(\rho^{(0)})^2\Psi^{(1)}
\end{equation}
\vspace{\baselineskip}
\newline
To proceed further, we need to define an inner product $\mathscr{I}\,:\,\mathscr{F}\rightarrow\mathbb{C}$ such that the operator $\hat{O}^P$ is self-adjoint with respect to it ($\mathscr{F}$ being the space of wavefunctions), i.e.:
\begin{equation}
    \mathscr{I}\left(\psi,\hat{O}^P\phi\right)=\mathscr{I}\left(\hat{O}^P\psi,\phi\right)
\end{equation}
The inner-product is defined as:
\begin{equation}
\label{eqn:Inner_Prod_Def}
    \mathscr{I}\left(\psi,\phi\right)=\int_{\mathscr{C}} \psi\,\phi\,dr_*
\end{equation}
The contour ($\mathscr{C}$) for integration is illustrated in Fig. \eqref{fig:Contour}. The details about the inner product definition and the choice of the contour are provided in Appendix \eqref{sec:Inner_Product_Defn}. The contour integral technique to compute corrections to QNM frequencies has been earlier for the Kerr-Newman black-hole case~\cite{Mark:2014aja}. However, to our knowledge, this is the first time this technique is used to compute the dCS corrections and the GR corrections in QNM frequencies for slow-rotating black-holes. 
Taking the inner product of the right-hand side and the left-hand side of \eqref{eqn:first_order_in_Chi_Polar} with $\Psi^{(0)}$, and using the above inner product definition, the first order (in $\chi$) correction to the QNM frequencies are:
\begin{equation}
\begin{split}
    \label{eqn:eigenFreq_correc}
    &\quad\rho^{(1)}=\frac{\mathscr{I}\left(\Psi^{(0)},\hat{V}^P\Psi^{(0)}\right)}{2\rho^{(0)}\,\mathscr{I}\left(\Psi^{(0)},\Psi^{(0)}\right)} \\&
    \implies \omega^{(1)}=-\frac{\mathscr{I}\left(\Psi^{(0)},\hat{V}^P\Psi^{(0)}\right)}{2\omega^{(0)}\,\mathscr{I}\left(\Psi^{(0)},\Psi^{(0)}\right)}
\end{split}
\end{equation}
We want to make the following remarks regarding the above result: 
First, even in the polar sector, we obtain dCS corrections (proportional to $\tilde{\alpha}^2$) to the QNM frequencies. This is the \emph{first significant difference} compared to the results in Ref. \cite{Wagle:2021tam}. Second, using Leaver's continued fractions method~\cite{Leaver:1985ax}, $\Psi^{(0)}$ and $\rho^{(0)}$ are known for the axial sector in the Schwarzschild background. Due to the isospectrality relation in GR, the zeroth-order frequencies $\rho^{(0)}$ for the polar sector are the same as the axial sector. Third, the zeroth-order eigenfunctions for the polar sector $\Psi^{(0)\mathrm{Z}}$ can be obtained from the eigenfunctions of the axial sector $\Psi^{(0)\mathrm{RW}}$ using the well known Chandrasekhar Transformation \cite{Chandrasekhar:1975nkd}. The expression for the polar wavefunction in terms of the axial wavefunction is given by:
 \begin{widetext}
 \begin{equation}
 \label{eqn:Chandra_Transf_Zer_From_RW}
\Psi_{l m \omega}^{(0)\,\mathrm{Z}}(r)=\frac{1}{\frac{4}{9} \lambda^{2}(\lambda+1)^{2}+4 M^{2} \omega^{2}}
\left[\left(\frac{2}{3} \lambda(\lambda+1)+\frac{6 M^{2}(r-2 M)}{r^{2}(\lambda r+3 M)}\right) \Psi_{l m \omega}^{(0) \mathrm{RW}}(r)+2 M \frac{d}{d r_{*}} \Psi_{l m \omega}^{(0)\mathrm{RW}}(r)\right]
\end{equation}
 \end{widetext}
where $\lambda=(l-1)(l+2)/2$. Thus, with the above inner product definition \eqref{eqn:Inner_Prod_Def}, we have all the ingredients to compute the correction in polar sector frequencies at linear order in spin.
\vspace{\baselineskip}
\newline
The above perturbation scheme is also useful in the axial sector. Ignoring the contribution of the pseudo-scalar field terms in the axial equation \eqref{eqn:Axial_Eq}, we can use the same technique discussed in the previous paragraph for the polar equation to compute some CS corrections to the QNM frequencies. 
Since we compute the corrections to the QNM frequencies only at the linear order in $\chi$, we can add the corrections in the QNM frequencies arising from the pseudo-scalar field terms in Eq.~\eqref{eqn:Axial_Eq} to the corrections due to the $\tilde{\alpha}^2$ terms to get the full correction at linear order in spin. In Ref. \cite{Wagle:2021tam} the authors calculated the corrections in QNM frequencies due to the  pseudo-scalar field terms by direct integration of equations. But they did not include the corrections due to the terms proportional to $\tilde{\alpha}^2$ in \eqref{eqn:Axial_Eq}. In this work, we have computed these corrections (that  were not included
in Ref.~\cite{Wagle:2021tam}) using the perturbation scheme discussed in this section. 

\subsection{The Results}
\label{sec:QNM_Results}
Having discussed the detailed procedure, in this subsection, we obtain the QNM frequencies for the first five fundamental ($n=0$) modes for the slowly rotating black-hole solution in the dCS theory. We set $\beta = 1$. As mentioned earlier, since the polar sector is decoupled, we can obtain the complete QNM frequencies. However, we need to solve the coupled differential equations for the axial sector to obtain the complete QNM frequencies. Hence, we only identify the corrections to the slow rotating Kerr QNM that were not included in Ref. \cite{Wagle:2021tam}. Since we are computing corrections at the linear order in spin, for the axial sector, we can add these corrections to the corrections reported in Ref. \cite{Wagle:2021tam} to get the full correction at linear order in spin. 

Although the procedure is suitable for any value of $l$, in this section, we will set $l = 2$. To compute the corrections to QNM frequencies, we use Eqs.~\eqref{eqn:eigenFreq_correc} for the Axial and Polar sector. To evaluate the contour integral 
[cf. Fig. \eqref{fig:Contour}] in Mathematica, we set the radius of the semicircular part of the contour as $\delta=0.2\,M$, continue the contour to $\text{Im}(r) = 200 M$ on both sides of the horizon point and approximate the wavefunctions by the first 80 terms in Leaver's continued fraction analysis \cite{Leaver:1985ax}. Note that the contour integral is independent of the choice of the radius of the semicircular arc of the contour, and the wavefunctions decay exponentially at the endpoints of the contour ($\text{Im}(r)=200\,M$). The analytical results reported in Table \ref{tab:Polar_Axial_Table} are robust to variation in $\delta$ and the choice of endpoints of the contour. More specifically, changing the value of end point of the contour from $\text{Im}(r) = 200\,M$ to $\text{Im}(r) = 400\,M$ changes the coefficients in Table \ref{tab:Polar_Axial_Table} only at the seventeenth decimal place. Varying $\delta$ in the range ($0.1\,M$, $0.4\,M$), the coefficients change only at the sixteenth decimal place. 
Similarly changing the number of terms used to approximate the wavefunctions in Leaver's analysis in the range of $80$ to $100$ changes the coefficients in Table \ref{tab:Polar_Axial_Table} only at the tenth decimal place. Thus, the results reported here are robust and do not 
depend on $\delta$ or the endpoints of the contour.
%
\vspace{\baselineskip}
\newline
The polar QNM frequencies can be expressed as:
\begin{equation}
\label{eqn:PolarQNM}
\omega^P_i= \omega_i^{\rm  GR}  +\, \frac{\tilde{\alpha}^2 m \chi}{M} \,c_i~~;~~
\omega_i^{\rm  GR} = \frac{1}{M} [a_i\,+\,m \chi\,b_i]
\end{equation}
where the coefficients $a_i$, $b_i$ and $c_i$ are tabulated in Table \ref{tab:Polar_Axial_Table}. 
We want to make the following remarks regarding the above result:
First, the coefficients $a_i$ match with the results in Ref. ~\cite{Kokkotas:1999bd} and the coefficients $b_i$ agree with the results in Ref.~\cite{Berti:2009kk}.
Second, the correction term proportional to only $m \chi$ is the correction due to spin in GR, whereas the correction term proportional to $\tilde{\alpha}^2 m \chi$ is the dCS correction. To get an estimate of the dCS corrections on the GR QNM frequencies, we need constraints on the dCS coupling parameter. In Refs. \cite{yanbei:2011fw,Yagi:2012ya}, the authors derived weak constraints on dCS coupling parameters from various astrophysical observations. A much more stringent constraint on dCS coupling parameter obtained using the measurements of the mass and the equatorial radius of the isolated neutron star PSR J0030+0451 by NICER (Neutron Star Interior Composition Explorer)~\cite{Riley_2019, Miller_2019, NICER10.1117/12.2231304} by Silva et al \cite{Silva:2020acr} (for $\beta = 1$) is:
%
%
\begin{equation}
\label{eqn:dCS_Coupling_Constraint0}
\sqrt{\alpha} \leqslant 8.5\, {\rm km} \, .
\end{equation}
This leads to:
\begin{equation}
\label{eqn:dCS_Coupling_Constraint}
\tilde{\alpha}  = \frac{\alpha}{M^2} 
\leq {33.11} 
\left(\frac{M_{\odot}}{M}\right)^2  \, .
\end{equation}
$\tilde{\alpha}$ is small for a $50 M_{\odot}$ black-hole and the corrections to the imaginary part is less than a percentage. Hence, binary black-hole collisions leading to $50 M_{\odot}$ black-holes can not provide any distinct signature. However, if the residual mass is around $15 M_{\odot}$ or less, like the BNS event~\cite{BNSevent:2017qsa}, then it can lead to a significant change in the QNM frequencies. For a $15 M_{\odot}$ black-hole, the above constraint \eqref{eqn:dCS_Coupling_Constraint} translates to:
\begin{eqnarray}
\tilde{\alpha}\leqslant 0.147
\end{eqnarray}
and is consistent with the small $\tilde{\alpha}$ approximation.
For $\tilde{\alpha}=0.1$, the ratio of the imaginary parts of the dCS correction to the purely GR correction in the first QNM frequency is:
\begin{eqnarray}
\text{Im}\left(\frac{\tilde{\alpha}^2c_1}{b_1}\right)\sim0.263
\end{eqnarray}
Thus, there is a significant correction in QNM frequency due to the dCS terms when compared to the GR correction and these corrections may potentially be observable in future missions~\cite{Evans:2016mbw}. {For $m\chi>0$, the dCS corrections make the magnitude of the imaginary part of the first quasi-normal mode less negative, hence decreasing the decay rate.}
%
%
\vspace{\baselineskip}
\newline
The third remark that we would like to make about our results is that although the contour-integration technique was used to compute corrections for the Kerr Newman black-hole in the past \cite{Mark:2014aja}, this is the first time the technique has been used to obtain QNM frequencies for modified gravity models. 
Lastly, due to the isospectrality theorem, GR spin correction is the same for both the axial and polar sectors~ \cite{Pani:2013pma}. As mentioned above, since the axial sector is coupled of the pseudo-scalar field ($\vartheta$), the axial QNM frequencies can be represented as:
\begin{equation}
\label{eqn:AxialQNM}
\omega^A_i = \omega_i^{\rm  GR} + 
\frac{\tilde{\alpha} ^2 m \, \chi}{M}\,   
d_i+\frac{\tilde{\alpha} ^2}{M}d^{\vartheta}_i(m,\chi) 
\end{equation}
where $d_i$ are the corrections that solely arise due to the corrections to the LHS in the Axial equation \eqref{eqn:Axial_Eq} and $d^{\vartheta}_i(m,\chi)$ are the corrections that arise due to the coupling to the pseudo-scalar field (RHS of the Eq. \eqref{eqn:Axial_Eq}). In Ref.~\cite{Wagle:2021tam}, the authors computed only $d^{\vartheta}_i$ terms and did not obtain the $d_i$ corrections. 
Here, we only obtain the corrections to the slow rotating Kerr QNM that were not included in Ref. \cite{Wagle:2021tam} and do not consider the corrections arising from the pseudo-scalar field $(\vartheta)$. Table \ref{tab:Polar_Axial_Table} contains the coefficients $d_i$. We plan to report the analytical evaluation of $d^{\vartheta}_i$ in a future publication. 
\section{Conclusions and Discussions}
\label{sec:Discussion}
In this work, we have analytically computed the fundamental mode ($n=0$) QNM frequencies for a slowly rotating black-hole solution in dCS gravity accurate to linear order in spin ($\chi)$ and quadratic order in the CS coupling parameter ($\alpha$). The corrections to the QNM frequencies reported here are at the same level of exactness as the background black-hole solution reported in Ref.~\cite{Yunes:2009hc}. In addition to Ref. \cite{Wagle:2021tam}, we found dCS corrections to the QNM frequencies in the polar sector. 
\begin{table*}[!htb]
{
\begin{tabular}{ |c|c|c|c|c|c| }
 \hline
 $i$ & 1 & 2 & 3 & 4 & 5 \\
 \hline 
 $a_i$ &  0.373672\,-0.0889623\,i & 0.346711\,-0.273915\,i & 0.301053\,-0.478277\,i & 0.251505\,-0.705148\,i & 0.207515\,-0.946845\,i  \\
 \hline
 $b_i$ & 0.0628831+0.000997934\,i & 0.07194+0.00638453\,i & 0.0860352+0.0224338\,i & 0.0957683+0.0508029\,i & 0.0971001+0.0866307\,i  \\
 \hline
 $c_i$ & -0.0489391+0.026246\,i & 0.259349\,-0.167831\,i & 0.703233\,-1.09307\,i & 0.969587\,-3.1162\,i & 0.527707\,-6.25558\,i  \\
 \hline
 $d_i$ & 0.0592739 - 0.189462\,i & -0.377947 - 0.504844\,i & -1.44408 - 0.434958\,i & -3.60286 + 1.11852\,i & -7.98051 + 7.12689\,i  \\
 \hline
\end{tabular}
}
 \caption{\label{tab:Polar_Axial_Table}
 QNM frequency coefficients for the polar sector $(a_i, b_i, c_i)$ and the axial sector $(a_i, b_i, d_i)$.}
\end{table*}
\vspace{\baselineskip}
\newline
To obtain the quasi-normal modes, we perturbed the background space-time linearly and computed the perturbation equations. 
As shown in Ref.~\cite{Wagle:2021tam}, the angular decomposed perturbation equations are invariant under the simultaneous transformations of axial, polar perturbation variables, $m$ and $a$. Due to this symmetry, the quasi-normal mode (QNM) frequencies are not affected by mode couplings (of $l$ to $l\pm 1$) at leading order in spin. This enormously simplified the perturbation equations, which we could combine in a specific manner. This leads to three Schroedinger-like equations --- one each for the scalar field, axial and polar sectors. The equations we have derived are again accurate to linear order in spin and quadratic order in the dCS coupling parameter. We find coupling between the dCS field and the axial sector metric perturbations, whereas the pseudo-scalar field does not couple to the polar sector metric perturbations. Interestingly, this feature was also found in spherically symmetric space-times~\cite{Bhattacharyya:2018hsj}.
\vspace{\baselineskip}
\newline
To compute the polar sector QNM frequencies, we analytically continue our mode-functions to the complex-$r$ plane. We defined an inner product using a contour integral along a contour that ensures the finiteness of the inner product. We employed a perturbation technique similar to non-degenerate perturbation theory in Quantum Mechanics and calculated the QNM frequencies in the polar sector. Again, we emphasize that our technique is similar to the perturbation theory in Quantum Mechanics, but our calculations are completely classical. We used the same technique to compute one part of the dCS corrections to QNM frequencies in the axial sector. 
Here, we only obtained the corrections to the slow rotating Kerr QNM that \emph{were not} included in Ref. \cite{Wagle:2021tam} and did not consider the corrections arising from the pseudo-scalar field $(\vartheta)$. The total dCS correction in the axial sector QNM frequencies is the sum of the two parts.
\vspace{\baselineskip}
\newline
{Our results shows that the dCS corrections are potentially observable when the final black-hole mass is less than 
$15 M_{\odot}$. Hence, the future BNS events~\cite{BNSevent:2017qsa} can potentially distinguish dCS and GR.
Specifically, assuming $\tilde{\alpha}=0.1$,
we found the ratio of the imaginary parts of the dCS correction to the purely GR correction in the first QNM frequency (for the polar sector) to be $0.263$. Also for $m\chi>0$, the dCS corrections make the magnitude of the imaginary part of the first QNM of the fundamental mode smaller, thereby decreasing the decay rate.} 
\vspace{\baselineskip}
\newline
Since only axial perturbations couple to the pseudo-scalar field. Our analysis shows that isospectrality between
odd and even parity perturbations are broken for the slowly rotating black-holes in dCS. Earlier, this was shown for the spherically symmetric black-holes~\cite{Bhattacharyya:2018hsj}. This leads to the question: Which term in the equations of motion \eqref{eqn:CS_EOM1} breaks the isospectrality relation for dCS?
To check this for the slowly rotating black-hole solution \eqref{eqn:Slow-rotMetric}, 
we ignored the C-Tensor term in the perturbation of equation \eqref{eqn:CS_EOM1} and $^*RR$ term in perturbation equation \eqref{eqn:CS_EOM2}. [Note that the rotating black-hole solution \eqref{eqn:Slow-rotMetric} is not a solution under these conditions.] In this case, the dCS corrections in QNM frequencies are isospectral for the axial and polar sectors. This confirms that the source of isospectrality breaking is the C-Tensor and $^*RR$ terms. 
\vspace{\baselineskip}
\newline
It seems that different strong-gravity corrections break the isospectrality differently. In the case of spherically symmetric space-times in $f(R)$ gravity, it was shown that the polar perturbations couple of the extra scalar gravitational perturbations while axial perturbations do not~\cite{Bhattacharyya:2017tyc,Bhattacharyya:2018qbe}. This is opposite to what we see in dCS gravity --- the axial perturbations couple of the extra scalar gravitational perturbations while polar perturbations do not. The future gravitational-wave detectors like the Cosmic Explorer~\cite{Evans:2016mbw} are expected to have a very high GW signal-to-noise ratio in the quasi-normal mode regime ${\rm SNR} > 50$~\cite{Nakano:2015uja}. Thus, the future detectors can help probe the QNM structure accurately and provide whether the strong-gravity violates parity or not!
\vspace{\baselineskip}
\newline
As we have mentioned earlier, we have not computed the QNM frequencies of the axial sector completely. This is because the axial perturbations are coupled to the pseudo-scalar field. The analytic perturbative technique developed in this work needs to be extended to accommodate the coupling with the pseudo-scalar field by modifying (or decoupling) the axial and the pseudo-scalar field equations using suitable operators and their commutators. This is currently under investigation.
%
\vspace{\baselineskip}
\newline
{We have analytically computed the QNM frequencies of a particular analytically obtained perturbative (slowly spinning-) solution of the dCS theory. In the literature, there have also been non-perturbative spinning solutions computed numerically for the dCS theory \cite{Delsate:2018ome}. For dCS black holes with general spins, metric perturbation does not have explicitly decoupled forms when $\alpha=0$, which means that solving for dCS QNM frequencies using metric perturbation will not be straightforward.  For nonlinear, scalarization scenarios discussed in Ref.~\cite{Gao:2018acg,Doneva:2021dcc}, it is plausible that, in a slowly-rotating special case, QNM frequencies can still be computed in the way discussed in this paper --- if a sufficiently simple closed-form analytical solution for the metric exists.  }


%
\section{Acknowledgements}
The authors thank Vitor Cardoso, Leonardo Gualtieri, Dongjun Li, Archana Pai, and Nicolas Yunes for helpful discussions. 
This work is part of the Dual Degree thesis project of MS. The work of SS is supported by SERB-MATRICS grant. 
Y.C.\ is supported by the Simons Foundation (Award Number 568762), the Brinson Foundation, and the National Science Foundation (Grants PHY--2011968, PHY--2011961 and PHY--1836809). 
The authors thank the service personnel in India and the USA whose untiring work allowed the authors to complete this work during the COVID-19 pandemic. 
\appendix
\section{Coefficient functions in Perturbation Equations}
\label{sec:Coeff_App}
%
%
In this section we list down all the potentials and the coefficient functions that occur in the Schroedinger-like equations in section \ref{sec:Schroedinger-like_Eq}. The coefficient functions are:
\begin{widetext}

\begin{equation}
f(r)=1-\frac{2M}{r} \, ;\qquad 
g(r)=\frac{6 i(l-1) l(l+1)(l+2) M^3}{r^{5} \omega \beta} \, ;\qquad
v(r)=-\frac{6 i M^3 \omega}{\kappa r^{5}}
\end{equation}
\begin{equation}
h(r)=-\frac{i\,M^2\left(r^{4} \omega^{2}\left(12(2 l(l+1)-1) M^{2}+15 M r+5 r^{2}\right)+144 M^{3}(2 M-r)\right)}{2 \beta r^{9} \omega^{2}} \, ; \qquad j(r)=\frac{72 i M^{5}(r-2 M)}{r^{8} \omega^{2}\beta} 
\end{equation}
\begin{equation}
n(r)=\frac{i\,M^2\left(-4224 M^{4}+3306 M^{3} r+48 M^{2} r^{2}\left(r^{2} \omega^{2}-15\right)+5 M r^{3}+15 r^{4}\right)}{4 \kappa l(l+1) r^{9}} \, ;\qquad
p(r)=\frac{12 i M^4(12 M-5 r)(2 M-r)}{\kappa l(l+1) r^{8}}
\end{equation}
The varius GR potentials are:
\begin{equation}
V_{\mathrm{eff}}^{S}=\left(1-\frac{2 M}{r}\right)\left[\frac{l(l+1)}{r^{2}}+\frac{2 M}{r^{3}}\right]+\operatorname{\chi m\omega} \frac{4 M^2}{r^{3}}
\end{equation}
\begin{equation}
V_{\mathrm{eff}}^{A}=\left(1-\frac{2 M}{r}\right)\left[\frac{l(l+1)}{r^{2}}-\frac{6 M}{r^{3}}+\frac{\chi m}{\omega} \frac{24 M^2(3 r-7 M)}{l(l+1) r^{6}}\right]+\chi m \omega \frac{4 M^2}{r^{3}}
\end{equation}
\begin{equation}
\begin{split}
V_{\mathrm{eff}}^{P}=&\chi m \omega \frac{4 M^2}{r^{3}}+ f(r)\left[\frac{2 M}{r^{3}}+\frac{(l-1)(l+2)}{3}\left(\frac{1}{r^{2}}+\frac{2(l-1)(l+2)(l(l+1)+1)}{(6 M+r(l(l+1)-2))^{2}}\right)\right.\\
&\left.+\frac{4 \chi m M^{2}}{r^{7} l(l+1)(6 M+r(l(l+1)-2))^{4} \omega}\left(27648 M^{6}+2592 M^{5} r(6 l(l+1)-19)\right.\right.\\
&\left.\left.+144 M^{4} r^{2}\left(230+l(l+1)(21 l(l+1)-148)+6 r^{2} \omega^{2}\right) \right.\right.\\
&\left.\left.+12 M^{2} r^{4}(l(l+1)-2)^{2}\left(l(l+1)(-12+5 l(l+1))+28 r^{2} \omega^{2}-4\right) \right.\right.\\
&\left.\left.+12 M^{3} r^{3}(l(l+1)-2)\left(374+l(l+1)(29 l(l+1)-200)+72 r^{2} \omega^{2}\right) \right.\right.\\
&\left.\left.+r^{6}(l(l+1)-2)^{3}\left(-3(l(l+1)-2)(l(l+1)+2)+2 r^{2}(l(l+1)-4) \omega^{2}\right) \right.\right.\\
&\left.\left.+M r^{5}(l(l+1)-2)^{2}((l(l+1)-2)(l(l+1)+2)(7 l(l+1)-38)\right.\right.\\
&\left.\left.\left.+24 r^{2}(2 l(l+1)-5) \omega^{2}\right)\right)\right]
\end{split}
\end{equation}
The $\Tilde{\alpha}^2$ correction functions in the potentials are:
\begin{equation}
    T_1(r)=\frac{3 (r-2 M)^2 \left(5 M^5 r^2+10 M^6 r+18 M^7\right)}{4 \beta  \kappa  r^{11} \omega }
\end{equation}
\begin{equation}
\begin{split}
     T_2(r)  =\frac{1}{56 \beta  \kappa  r^{12} \omega }& \left(2016 l (l+1) M^6 r^3 \omega  (r-2 M)-m \chi  \left(2016 l^2 M^7 (r-2 M)^2+2016 l M^7 (r-2 M)^2\right.\right.\\
        &\left.\left.-8064 M^8 r^3 \omega ^2+120 M^6 r^5 \omega ^2+21 M^5 r^4 \left(201 M^2 \omega ^2+10\right)+70 M^5 r^6 \omega ^2-4284 M^7 r^2\right.\right.\\
        &\left.\left.+10416 M^8 r-7056 M^9\right)\right)
\end{split}
\end{equation}
\begin{equation}
    \begin{split}
        A_1(r)=\frac{1}{4 \beta  \kappa  l (l+1) r^{12} \omega}&\left((r-2 M)^2 \left(-15 M^4 r^4 \left(5 l^2+5 l-4 \left(6 M^2 \omega ^2+5\right)\right)-40 \left(l^2+l+7\right) M^6 r^2\right.\right.\\
        &\left.\left.-15 \left(3 l^2+3 l+4\right) M^5 r^3+18 \left(83 l^2+83 l-640\right) M^7 r+120 M^5 r^5 \omega ^2+30 M^4 r^6 \omega ^2+19440 M^8\right)\right)
    \end{split}
\end{equation}
\begin{equation}
    \begin{split}
        A_2(r)=-\frac{1}{56 \beta  \kappa  l (l+1) r^{13} \omega}&\left(1680 M^6 r^4 \left(2 l^2+2 l-78 M^2 \omega ^2+3\right)+21 M^5 r^5 \left(l^2 \left(9 M^2 \omega ^2-160\right)+l \left(9 M^2 \omega ^2-160\right)\right.\right.\\
        &\left.\left.+20 \left(212 M^2 \omega ^2-9\right)\right)+70 \left(l^2+l-18\right) M^5 r^7 \omega ^2+30 M^4 r^6 \left(l^2 \left(4 M^2 \omega ^2+35\right)\right.\right.\\
        &\left.\left.+l \left(4 M^2 \omega ^2+35\right)-84 M^2 \omega ^2\right)+168 \left(48 l^4+96 l^3+1141 l^2+1093 l+1966\right) M^8 r^2\right.\\
        &\left.-168 \left(12 l^4+24 l^3+303 l^2+291 l+11\right) M^7 r^3-1008 \left(8 l^4+16 l^3+185 l^2+177 l+1168\right) M^9 r\right.\\
        &\left.-1680 M^4 r^8 \omega ^2+1088640 M^{10}\right)
    \end{split}
\end{equation}
\begin{equation}
    \begin{split}
        P_1(r)=&-\frac{1}{56 \beta  \kappa  l (l+1) r^{12} \omega  \left(\left(l^2+l-2\right) r+6 M\right)^3}\left(M^3 (r-2 M)^2 \left(-12 M^4 r^4 \left(6479 l^6+19437 l^5-17072 l^4-66539 l^3\right.\right.\right.\\
        &\left.\left.\left.+57739 l^2+94248 l-5670 M^2 \omega ^2-150992\right)-6 M^3 r^5 \left(-42 l^2 \left(99 M^2 \omega ^2+1555\right)+315 l^6+945 l^5+5635 l^4+9695 l^3\right.\right.\right.\\
        &\left.\left.\left.-14 l \left(297 M^2 \omega ^2+5000\right)+4 \left(639 M^2 \omega ^2+29680\right)\right)+560 \left(l^2+l-2\right)^2 M^2 r^8 \omega ^2-18 M^2 r^6 \left(-7 l^4 \left(18 M^2 \omega ^2+355\right)\right.\right.\right.\\
        &\left.\left.\left.-7 l^3 \left(36 M^2 \omega ^2+785\right)+l^2 \left(7140-342 M^2 \omega ^2\right)+105 l^6+315 l^5+l \left(9940-216 M^2 \omega ^2\right)+96 M^2 \omega ^2-9520\right)\right.\right.\\
        &\left.\left.+60 \left(l^2+l-2\right) M r^7 \left(5 l^2 \left(4 M^2 \omega ^2-21\right)+35 l^4+70 l^3+20 l \left(M^2 \omega ^2-7\right)+58 M^2 \omega ^2+140\right)\right.\right.\\
        &\left.\left.+18 \left(147273 l^4+294546 l^3-960469 l^2-1107742 l+1603992\right) M^6 r^2+27 \left(5439 l^6+16317 l^5-68265 l^4\right.\right.\right.\\
        &\left.\left.\left.-163725 l^3+181898 l^2+266480 l-267264\right) M^5 r^3+432 \left(40551 l^2+40551 l-130246\right) M^7 r+39680928 M^8\right)\right)\\&
        +\frac{15\omega(48 M^8-32 M^7 r + M^4 r^4)}{2l(l+1)\beta\kappa r^8}
    \end{split}
\end{equation}
\begin{equation}
    \begin{split}
        P_2(r)=&-\frac{1}{56 l (l+1) r^{13} \left(6 M+\left(l^2+l-2\right) r\right)^4 \beta  \kappa  \omega }\left(952342272 M^{14}+10368 \left(40551 l^2+40551 l-222100\right) r M^{13}\right.\\
        &\left.+1728 \left(22344 l^4+44688 l^3-468949 l^2-491293 l+1320255\right) r^2 M^{12}-216 \left(38955 l^6+116865 l^5+327754 l^4 \right.\right.\\
        &\left.\left.+460733 l^3-2836929 l^2-3047818 l+5572728\right) r^3 M^{11}-36 r^4 \left(51324 l^8+205296 l^7-181881 l^6-1264179 l^5\right.\right.\\
        &\left.\left.-1568026 l^4-789575 l^3+6658635 l^2+6808510 l+984312 M^2 \omega ^2-10830664\right) M^{10}-6 r^5 \left(15351 l^{10}+76755 l^9\right.\right.\\
        &\left.\left.-274595 l^8-1558910 l^7-277387 l^6+4946395 l^5+4242251 l^4-1731728 l^3+4 \left(699678 \omega ^2 M^2-2652053\right) l^2\right.\right.\\
        &\left.\left.+72 \left(38871 \omega ^2 M^2-119932\right) l-9312408 \omega ^2 M^2+17313024\right) M^9+3 \left(29999 l^{10}+149995 l^9-81007 l^8\right.\right.\\
        &\left.\left.-1223998 l^7-614347 l^6+3070931 l^5+\left(2263351-707616 \omega ^2 M^2\right) l^4-48 \left(29484 M^2 \omega ^2+48323\right) l^3\right.\right.\\
        &\left.\left.+\left(5889888 \omega ^2 M^2-6264220\right) l^2+32 \left(206172 \omega ^2 M^2-135739\right) l+48 \left(222776-212043 \omega ^2 M^2\right)\right) r^6 M^8\right.\\
        &\left.-4 r^7 \left(5108 l^{10}+25540 l^9-3659 l^8-167876 l^7+\left(4498-10962 \omega ^2 M^2\right) l^6+\left(708328-32886 \omega ^2 M^2\right) l^5\right.\right.\\
        &\left.\left.+\left(397141-407718 \omega ^2 M^2\right) l^4-2 \left(380313 M^2 \omega ^2+316600\right) l^3+22 \left(52965 \omega ^2 M^2-68264\right) l^2\right.\right.\\
        &\left.\left.+2 \left(770031 \omega ^2 M^2-515756\right) l+180 \left(12208-9477 \omega ^2 M^2\right)\right) M^7+2 \left(1505 l^{10}+7525 l^9+35 \left(243 \omega ^2 M^2-1033\right) l^8\right.\right.\\
        &\left.\left.+70 \left(486 \omega ^2 M^2-2711\right) l^7+\left(14118 M^2 \omega ^2+9205\right) l^6+\left(723415-76716 \omega ^2 M^2\right) l^5-7 \left(5409 \omega ^2 M^2-55015\right) l^4\right.\right.\\
        &\left.\left.+\left(91824 \omega ^2 M^2-671930\right) l^3+20 \left(9396 \omega ^2 M^2-35875\right) l^2+8 \left(15444 \omega ^2 M^2-28385\right) l-798960 \omega ^2 M^2\right.\right.\\
        &\left.\left.+715680\right) r^8 M^6+3 \left(l^2+l-2\right) \left(7 \left(9 \omega ^2 M^2-190\right) l^8+28 \left(9 \omega ^2 M^2-190\right) l^7+\left(7140-1039 \omega ^2 M^2\right) l^6\right.\right.\\
        &\left.\left.+\left(40040-3999 \omega ^2 M^2\right) l^5+\left(7446 M^2 \omega ^2+3710\right) l^4+\left(21851 \omega ^2 M^2-65520\right) l^3-2 \left(1037 M^2 \omega ^2+8960\right) l^2\right.\right.\\
        &\left.\left.-4 \left(3427 \omega ^2 M^2-5600\right) l+56 \left(300-2857 \omega ^2 M^2\right)\right) r^9 M^5+70 \left(l^2+l-2\right)^3 \left(l^4+2 l^3-10 l^2-11 l+28\right) r^{11} \omega ^2 M^5\right.\\
        &\left.+10 \left(l^2+l-2\right)^2 \left(3 \left(4 M^2 \omega ^2+35\right) l^6+9 \left(4 M^2 \omega ^2+35\right) l^5+3 \left(34 \omega ^2 M^2-35\right) l^4+3 \left(48 \omega ^2 M^2-245\right) l^3\right.\right.\\
        &\left.\left.-518 \omega ^2 M^2 l^2+\left(420-584 \omega ^2 M^2\right) l-5324 M^2 \omega ^2\right) r^{10} M^4+420 \left(l^2+l-2\right)^4 r^{12} \omega ^2 M^4\right)\\&
        -\frac{15 M^4 \omega (6M^2 + 3(l^2+l-2)Mr-(l^2+l-2)r^2)(24M^3-4M^2r-2Mr^2-r^3)}{2l(l+1)r^9(6M+(l^2+l-2)r)\beta\kappa}
    \end{split}
\end{equation}
\end{widetext}
\section{Eliminating Angular Functions from Perturbation Equations and Orthogonality Properties}
\label{sec:X_cos_W_cos}
Kojima like decomposition highlighted in Section \ref{sec:Kojima_Decomp} gives us equations that contain both angular and radial functions summed up over all $l$ and $m$ modes. As mentioned in Section \ref{sec:Kojima_Decomp}, to eliminate the angular functions and to get radial equations in individual $l$, $m$ modes, we need 
    to use the perturbation equations along with some orthogonality properties of spherical harmonics. For example, we can multiply the perturbation equations \eqref{eqn:Eq_Group1} by $Y^{*}_{l m}$ and integrate over the solid angle to get rid of the angular functions. The detailed procedure for getting rid of the angular functions from all the 11 perturbation equations has been highlighted in \cite{Pani:2013pma} for the GR case. The only point of difference from the GR analysis for the case of the CS slowly rotating background are the $\tilde{s}_{l m}$ and $\tilde{t}_{l m}$ terms in \eqref{eqn:Eq_Group3_1} and \eqref{eqn:Eq_Group3_2}. These terms vanish for a slowly rotating Kerr background. In this section, we demonstrate one of the procedures to deal with these terms. 
\vspace{\baselineskip}
\newline
We first mention some of the important relations, definitions and orthogonality properties that we will be needing. The scalar product definition is:
\begin{equation}
\label{eqn:ScalarProduct}
\langle f, g \rangle\,\equiv \int d \Omega f^{*} g=\int d \theta d \varphi \sin \theta f^{*} g
\end{equation}
The scalar spherical harmonics are governed by the following equation:
\begin{equation}
   Y^{l m}_{,\theta\theta}+\cot\theta Y^{l m}_{,\theta}+\frac{1}{\sin^2\theta}Y^{l m}_{,\varphi\varphi}+l(l+1)Y^{l m}=0 
\end{equation}
and are subject to the orthogonality relation:
\begin{equation}
\langle Y^{l m}, Y^{l^{\prime} m^{\prime}}\rangle \,=\delta^{l l^{\prime}} \delta^{m m^{\prime}}
\end{equation}
Some relevant recursion properties of scalar spherical harmonics are:
\begin{equation}
\label{eqn:cosIntoYlm}
\cos \theta Y^{l m}=\mathcal{Q}_{l+1\,m} Y^{l+1\,m}+\mathcal{Q}_{l m} Y^{l-1\,m}
\end{equation}
\begin{equation}
\sin \theta Y_{, \theta}^{l m}=\mathcal{Q}_{l+1\,m} l Y^{l+1\,m}-\mathcal{Q}_{l m}(l+1\,m) Y^{l-1\,m}
\end{equation}
Some other important integral relations among $Y^{l m}$, $W^{l m}$, $X^{l m}$ are:
\begin{equation}
\label{eqn:WX_int1}
\int d \Omega\left[W^{* l^{\prime} m^{\prime}} Y_{, \varphi}^{l m}-X^{* l^{\prime} m^{\prime}} Y_{, \theta}^{l m}\right]=i m(l(l+1)-2) \delta_{m m^{\prime}} \delta_{l l^{\prime}}
\end{equation}
\begin{equation}
\label{eqn:WX_int2}
\int d \Omega \cos \theta\left[W^{* l m} W^{l m}+\frac{X^{* l m} X^{l m}}{\sin \theta^{2}}\right]=0
\end{equation}
Note that the above integral \eqref{eqn:WX_int2} is zero for same ($l\,m$) indices on all the functions. A more general equation involves $l$ coupled to $l^{\prime}\pm 1$, $l^{\prime}\pm 3$..... Another important integral is:
\begin{equation}
\label{eqn:WX_int3}
\int d \Omega\left[\frac{W^{* l^{\prime} m^{\prime}} X^{l m}-X^{* l^{\prime} m^{\prime}} W^{l m}}{\sin \theta}\right]=0
\end{equation}
Also define the following operator:
\begin{equation}
\label{eqn:LOperator}
\mathcal{L}_{2}^{\pm 1} N_{l m} \equiv-(l+1) \mathcal{Q}_{l m} N_{l-1 m}+l \mathcal{Q}_{l+1 m} N_{l+1 m}
\end{equation}
where
\begin{equation}
\label{eqn:Qlm}
\mathcal{Q}_{l m}=\sqrt{\frac{l^{2}-m^{2}}{4 l^{2}-1}}
\end{equation}
The above operator is useful in describing the following integral:
\begin{widetext}
\begin{equation}
\label{eqn:WX_int_L}
\sum_{l, m}\left\{\frac{g_{l m}}{(l(l+1)-2)}\int d \Omega \left[\sin\theta W^{* l^{\prime} m^{\prime}} Y_{,\theta}^{l m}+\frac{X^{* l^{\prime} m^{\prime}} Y_{,\varphi}^{l m}}{\sin \theta}\right]\right\}=\mathcal{L}_{2}^{\pm 1} g_{l^{\prime} m^{\prime}}
\end{equation}
\end{widetext}
We will also need some properties of the Spin weighted spherical harmonics. For more details, please refer \cite{Goldberg:1966uu}. Spin weighted spherical harmonics can be obtained from the usual ($s=0$) scalar spherical harmonics $Y^{l m}$ by using ladder operators:
\begin{equation}
{ }_{s} Y_{l m}=\left\{\begin{array}{ll}
\sqrt{\frac{(l-s) !}{(l+s) !}} \eth^{s} Y_{l m}, & 0 \leq s \leq l \\
\sqrt{\frac{(l+s) !}{(l-s) !}}(-1)^{s} \overline{\eth}^{-s} Y_{l m}, & -l \leq s \leq 0 ; \\
0, & l<|s| .
\end{array}\right.
\end{equation}
where single raising and lowering operations are:
\begin{equation}
\label{eqn:raising_action}
\eth\left({ }_{s} Y_{l m}\right)=+\sqrt{(l-s)(l+s+1)} \,{}_{s+1} Y_{l m}
\end{equation}
\begin{equation}
\label{eqn:lowering_action}
\overline{\eth}\left({ }_{s} Y_{l m}\right)=-\sqrt{(l+s)(l-s+1)} \,{}_{s-1} Y_{l m}
\end{equation}
and the ladder operators are defined as:
\begin{equation}
\label{eqn:raising_operator}
\eth \eta=-(\sin \theta)^{s}\left[\frac{\partial}{\partial \theta}+i \csc \theta \frac{\partial}{\partial \varphi}\right](\sin \theta)^{-s} \eta
\end{equation}
\begin{equation}
\label{eqn:lowering_operator}
\overline{\eth} \eta=-(\sin \theta)^{-s}\left[\frac{\partial}{\partial \theta}-i \csc \theta \frac{\partial}{\partial \varphi}\right](\sin \theta)^{s} \eta
\end{equation}
with $\eta$ being a spin weight $s$ field. The orthogonality of the spin-weighted spherical harmonics is:
\begin{equation}
   \langle {}_{-2}Y_{l^{\prime}m^{\prime}}\,,\,{}_{-2}Y_{l m}\rangle =\delta_{l l^{\prime}}\,\delta_{m m^{\prime}}
\end{equation}
The forms of some of the relevant spin-weighted spherical harmonics are:
\begin{equation}
\label{eqn:S-2}
{}_{-2} Y_{l m}(\theta, \varphi) \equiv \frac{W_{l m}(\theta, \varphi)-i X_{l m}(\theta, \varphi) / \sin \theta}{\sqrt{l(l+1)(l(l+1)-2)}}
\end{equation}
\begin{equation}
    \label{eqn:S-1}
    {}_{-1} Y_{l m}(\theta, \varphi) \equiv\frac{-1}{\sqrt{l(l+1)}}\left(Y_{l m,\theta}-i\frac{Y_{l m,\varphi}}{\sin \theta}\right)
\end{equation}
Now we highlight the procedure to deal with the $\tilde{s}_{l m}$ and $\tilde{t}_{l m}$ terms in \eqref{eqn:Eq_Group3_1} and \eqref{eqn:Eq_Group3_2}. Following \cite{Pani:2013pma}, let us call equation \eqref{eqn:Eq_Group3_1} multiplied by $\sin\theta$ as $\delta\varepsilon_{\theta\varphi}$ and \eqref{eqn:Eq_Group3_2} multiplied by $\sin\theta$ as $\delta\varepsilon_{(-)}$. Then we have:
\begin{widetext}
\begin{equation}
\label{eqn:Group3_1_int}
\begin{split}
   0= &\int\frac{d\Omega}{(l(l +1)-2)} \left(W^{*}_{ l^{\prime} m^{\prime}}\delta\varepsilon_{(-)}+\frac{X^{*}_{ l^{\prime} m^{\prime}}}{\sin\theta}\delta\varepsilon_{\theta\varphi}\right)\sim\\&-im^{\prime}\,f_{ l^{\prime} m^{\prime}}+l^{\prime}(l^{\prime} +1)s_{ l^{\prime} m^{\prime}}+\mathcal{L}_{2}^{\pm 1} g_{l^{\prime} m^{\prime}}+\sum_{l, m}\left\{\frac{\tilde{t}_{l m}}{(l(l+1)-2)}\int d \Omega \cos\theta \left[\frac{-W^{* l^{\prime} m^{\prime}} X^{l m}+X^{* l^{\prime} m^{\prime}} W^{l m}}{\sin \theta}\right]\right\}
\end{split}
\end{equation}
Similarly
\begin{equation}
\label{eqn:Group3_2_int}
\begin{split}
   0= &\int\frac{d\Omega}{(l(l +1)-2)} \left(W^{*}_{ l^{\prime} m^{\prime}}\delta\varepsilon_{\theta\varphi}-\frac{X^{*}_{ l^{\prime} m^{\prime}}}{\sin\theta}\delta\varepsilon_{(-)}\right)\sim\\& im^{\prime}\,g_{ l^{\prime} m^{\prime}}+l^{\prime}(l^{\prime} +1)t_{ l^{\prime} m^{\prime}}+\mathcal{L}_{2}^{\pm 1} g_{l^{\prime} m^{\prime}}+\sum_{l, m}\left\{\frac{-\tilde{s}_{l m}}{(l(l+1)-2)}\int d \Omega \cos\theta \left[\frac{-W^{* l^{\prime} m^{\prime}} X^{l m}+X^{* l^{\prime} m^{\prime}} W^{l m}}{\sin \theta}\right]\right\}
\end{split}
\end{equation}
\end{widetext}
In the above 2 equations we have already integrated the terms which are also present in the GR case, but have kept the extra integral pieces that arise in the dCS case as it is. Note that any $\delta_{l^{\prime}l\pm 1}$ and $\mathcal{L}_{2}^{\pm 1}$ terms are going to lead to couplings of $l$-mode terms with $(l\pm 1)$-modes. As discussed in Section \ref{sec:Simplifying_Pert_Eq} of the paper, these terms can be neglected for the purposes of QNM frequency calculation. In this appendix, in some steps some of these $\delta_{l^{\prime}l\pm 1}$ and $\mathcal{L}_{2}^{\pm 1}$ terms are explicitly mentioned, but they are of no consequence for QNM frequency calculation. Note that we have neglected the integral with $\tilde{s}$ in \eqref{eqn:Group3_1_int} and the integral with $\tilde{t}$ in \eqref{eqn:Group3_2_int} because for the relevant case of $l^{\prime}=l$, they vanish due to the equation \eqref{eqn:WX_int2}. What we are left with is evaluating the integral of the form:
\begin{widetext}
\begin{equation}
    \mathcal{I}=\sum_{l, m}\left\{\frac{\tilde{N}_{l m}}{(l(l+1)-2)}\int d \Omega \cos\theta \left[\frac{-W^{* l^{\prime} m^{\prime}} X^{l m}+X^{* l^{\prime} m^{\prime}} W^{l m}}{\sin \theta}\right]\right\}
\end{equation}
To evaluate this integral, we observe that $\mathcal{I}$ is related to the scalar product of spin-2 spherical harmonics a
\begin{equation}
\label{eqn:extra_integral_term}
    \mathcal{I}=-i\sum_{l, m}\tilde{N}_{l m}\frac{\sqrt{l(l+1)(l(l+1)-2)l^{\prime}(l^{\prime}+1)(l^{\prime}(l^{\prime}+1)-2)}}{(l(l+1)-2)}\langle {}_{-2}Y_{l^{\prime}m^{\prime}}\,,\,\cos\theta{}_{-2}Y_{l m}\rangle
\end{equation}
Thus the problem reduces to calculating the scalar product of the spin-2 spherical harmonics in the above equation. For this we operate equation \eqref{eqn:cosIntoYlm} by the lowering operator ($\overline{\eth}$) \eqref{eqn:lowering_operator} for spin weight 0 to get:
\begin{equation}
    -\sin\theta Y_{l m}+l(l+1)\cos\theta {}_{-1}Y_{l m}=\mathcal{Q}_{l+1\,m}\sqrt{(l+1)(l+2)}{}_{-1}Y_{l+1\, m}+\mathcal{Q}_{l m}\sqrt{(l-1)l}{}_{-1}Y_{l-1\, m}
\end{equation}
We then apply the lowering operator for spin weight 1 to the above equation to get:
\begin{equation}
    \cos\theta {}_{-2}Y_{l m}=-\mathcal{Q}_{l+1\,m}\sqrt{\frac{(l +3)}{(l-1)l(l+1)}}{}_{-2}Y_{l+1\, m}-\mathcal{Q}_{l m}\sqrt{\frac{(l -2)}{l(l+1)(l+2)}}{}_{-2}Y_{l-1\, m}-\frac{2\sin\theta}{\sqrt{l(l+1)-2}}{}_{-1}Y_{l m}
\end{equation}
Taking scalar product of the above equation with ${}_{-2}Y_{l^{\prime}m^{\prime}}$, we get:
\begin{equation}
\label{eqn:s-2cosS-2_int}
\begin{split}
    \langle {}_{-2}Y_{l^{\prime}m^{\prime}}\,,\,\cos\theta{}_{-2}Y_{l m}\rangle =&-\mathcal{Q}_{l+1\,m}\sqrt{\frac{(l +3)}{(l-1)l(l+1)}}\delta_{m m^{\prime}}\delta_{l+1\,l^{\prime}}-\mathcal{Q}_{l m}\sqrt{\frac{(l -2)}{l(l+1)(l+2)}}\delta_{m m^{\prime}}\delta_{l-1\,l^{\prime}}\\&
    -\frac{2}{\sqrt{l(l+1)-2}} \langle {}_{-2}Y_{l^{\prime}m^{\prime}}\,,\,\sin\theta{}_{-1}Y_{l m}\rangle
\end{split}
\end{equation}
We have now reduced the problem to evaluating $<{}_{-2}Y_{l^{\prime}m^{\prime}}\,,\,\sin\theta{}_{-1}Y_{l m}>$. This can be evaluated directly by substituting the harmonics from equations \eqref{eqn:S-2} and \eqref{eqn:S-1} to get:
\begin{equation}
\label{eqn:S-1S-2_int}
\begin{split}
    \langle {}_{-2}Y_{l^{\prime}m^{\prime}}\,,\,\sin\theta{}_{-1}Y_{l m}\rangle =&-\frac{m\sqrt{(l(l+1)-2)}}{l(l+1)}\delta_{m m^{\prime}}\delta_{l l^{\prime}}\\&
    -\frac{1}{\sqrt{l(l+1)l^{\prime}(l^{\prime}+1)(l^{\prime}(l^{\prime}+1)-2)}}\int d \Omega \left[\sin\theta W^{* l^{\prime} m^{\prime}} Y_{,\theta}^{l m}+\frac{X^{* l^{\prime} m^{\prime}} Y_{,\varphi}^{l m}}{\sin \theta}\right]
\end{split}
\end{equation}
\end{widetext}
The un-evaluated integral above is directly related to the operator $\mathcal{L}_{2}^{\pm 1}$ (see Eq. \eqref{eqn:WX_int_L}). 
\vspace{\baselineskip}
\newline
Thus we now have all the terms required to evaluate the extra integral term $\mathcal{I}$ defined in equation \eqref{eqn:extra_integral_term}. Note that all the $\delta_{l^{\prime}l\pm 1}$ and $\mathcal{L}_{2}^{\pm 1}$ terms are going to lead to couplings of the $l$-mode terms with $(l\pm 1)$-modes. As discussed in Section \ref{sec:Simplifying_Pert_Eq} of the paper, these terms can be neglected for the purposes of QNM frequency calculations. Therefore, we only require the following scalar product relation (which we get by using Eq. \eqref{eqn:S-1S-2_int} in Eq. \eqref{eqn:s-2cosS-2_int} and setting $l^{\prime}=l$):
\begin{equation}
\begin{split}
\langle{}_{-2}Y_{l m}\,,\,\cos\theta{}_{-2}Y_{l m}\rangle \,&=\frac{2m}{l(l+1)}\quad\quad\text{for} \quad l\geqslant2\\&
    =0\qquad \qquad \text{otherwise}
\end{split}    
\end{equation}
Ignoring the mode coupling terms in the extra integral $\mathcal{I}$, we get:
\begin{equation}
    \mathcal{I}=-2im\tilde{N}_{l^{\prime} m^{\prime}}
\end{equation}
The equations \eqref{eqn:Group3_1_int}, ripped of all the angular functions and keeping only the terms relevant for QNM frequency calculations, finally take the following form after replacing all the 'primed' indices by 'un-primed' ones:
\begin{equation}
    -im\,f_{ l m}+l(l +1)\,s_{ l m}-2im\,\tilde{t}_{l m}=0
\end{equation}
\begin{equation}
    im\,g_{ l m}+l(l +1)\,t_{ l m}+2im\,\tilde{s}_{l m}=0
\end{equation}
Note that the extra $\tilde{t}_{l m}$ and $\tilde{s}_{l m}$ terms will not be present for $l\leqslant 1$ because the spin-2 spherical harmonics vanish for such cases. 
\section{Inner Product Definition: Analytic Continuation to the Complex Plane}
\label{sec:Inner_Product_Defn}
We need to define an inner product such that the operator $\frac{d^2}{dr_*^2}$ is self-adjoint. This would, in turn, result in any operator having the same form as $\hat{O}^P$ being self-adjoint. QNM wavefunctions are not square integrable along the real r-axis, since they diverge close to the horizon and at spatial infinity ($r\rightarrow\infty$). Therefore the usual inner product candidates defined using integrals of wavefunctions will not be convergent. A way around this issue, as highlighted in Ref. \cite{Mark:2014aja}, is to analytically continue the QNM wavefunctions to the complex plane. In order to get a finite inner product, we observe that the outgoing boundary condition implies that the QNM wavefunctions are purely outgoing (~ $e^{i\omega r_*}$) at spatial infinity. Therefore the wavefunctions decay exponentially as $r\rightarrow i\, \infty$ (for $\text{Re}(\omega)>0$ - which is true for the QNM frequencies). Thus we can get a finite inner product of two wavefunctions as:
\begin{equation}
\label{eqn:Inner_Prod_Def01}
    \mathscr{I}\left(\psi,\phi\right)=\int_{\mathscr{C}} \psi\,\phi\,dr_*
\end{equation}
where $\mathscr{C}$ is a contour (Fig. \ref{fig:Contour}) that begins to the right of the horizon ($r=2M$) position at positive imaginary infinity, runs down parallel to the imaginary axis, encircles the the horizon point and goes back to positive infinity, this time on the left of the horizon point. For ease of calculation, we keep the contour symmetric about the horizon point ($r=2M$). Note that the contour integral of two wavefunctions will not be zero because the wavefunctions are not analytic in the region enclosed by $\mathscr{C}$. They blow up at the horizon point and thus require a branch cut that runs parallel to the imaginary axis emanating from $r=2M$. It can be easily verified that the operator $\frac{d^2}{dr_*^2}$ (equivalently $\hat{O}^P$) is self-adjoint with respect to the inner product \eqref{eqn:Inner_Prod_Def}. Thus, we can calculate the QNM corrections mentioned in \eqref{eqn:eigenFreq_correc} simply by evaluating integrals along the contour ($\mathscr{C}$) in the complex-r plane.
\section{Fermi Estimate Analysis}
\label{app:Fermi_Estimate}
Fermi estimate is a quick way to obtain $\omega$ dependence on the CS coupling parameter ($\alpha$). {It has been described in \cite{Wagle:2021tam}}. In Fermi estimate, we replace any radial derivative by a characteristic length scale $\mathcal{R}$, i.e., $\partial_r\rightarrow \mathcal{R}^{-1}$,  in the relevant equations to make them algebraic instead of differential. We then evaluate the algebraic equation at the characteristic length scale. Solving the algebraic equations for $\omega$ provides a way to obtain $\omega$ dependence on certain parameters of the theory. 

Let us first apply Fermi estimate to the polar equation \eqref{eqn:Polar_Eq}. Replacing the derivatives in \eqref{eqn:Polar_Eq} gives:
\begin{widetext}
\begin{equation}
\left[G(\mathcal{R})+\Tilde{\alpha}^2 \chi\,m\, G_P(\mathcal{R})+\omega^{2}-(V_{\mathrm{eff}}^{P}(\mathcal{R}, \chi)+\Tilde{\alpha}^2\,\chi\,m\,P_2(\mathcal{R}))\right] \Psi^{\mathrm{Z}}_{l m}=0
\end{equation}
From the equation above, it is clear that, when expanded perturbatively in powers of $\alpha$, $\omega_{\rm Polar}$ does depend on the CS coupling parameter. This is different from what is claimed in Ref.~\cite{Wagle:2021tam}. Taking into account that $V_{\mathrm{eff}}^{P}$ is linear in $\omega$, it is easy to see that the correction in $\omega$ from the GR case will be at order $\tilde{\alpha}^2$ for small $\tilde{\alpha}$. 
\vspace{\baselineskip}
\newline
Let us now apply Fermi estimate to the Axial \eqref{eqn:Axial_Eq} and $\Theta$ \eqref{eqn:Theta_Eq} sector. This leads to the following equations:
\begin{eqnarray}
\left[G(\mathcal{R})+\Tilde{\alpha}^2 \chi\,m\, G_S(\mathcal{R})+\omega^{2}-V_{\mathrm{eff}}^{Smod}(\mathcal{R})\right] R_{l m} &=& \Tilde{\alpha} H(\mathcal{R}) \Psi^{\mathrm{RW}}_{l m} \\
\left[G(\mathcal{R})+\Tilde{\alpha}^2\chi\,m\, G_A(\mathcal{R})+\omega^{2}-V_{\mathrm{eff}}^{Amod}(\mathcal{R})\right] \Psi^{\mathrm{RW}}_{l m} &=& 
\Tilde{\alpha} I(\mathcal{R}) R_{l m}
\end{eqnarray}
Multiplying the above two equations and cancelling of the eigenfunctions, we get:
\begin{equation}
\left[G+\Tilde{\alpha}^2\chi\,m\, G_S+\omega^{2}-V_{\text {eff }}^{Smod}\right]\left[G+\Tilde{\alpha}^2 \chi\,m\, G_A+\omega^{2}-V_{\text {eff }}^{Amod}\right]=\Tilde{\alpha}^{2} I H
\end{equation}
This is a quadratic equation in $\omega^2$, solving for small $\Tilde{\alpha}$, we get:
\begin{equation}
    \omega^2=\left(V_{\mathrm{eff}}^{A}-G\right) + \Tilde{\alpha}^2\left(\chi\,m\,(A_2-G_A)+\frac{I H}{\left(V_{\text {eff }}^{A}-V_{\text {eff }}^{S}\right)} \right)
\end{equation}
\end{widetext}
Thus $\omega$ can be expressed as:
\begin{equation}
\omega=\omega_{\mathrm{GR}}+\Tilde{\alpha}^2 \delta \omega+\mathcal{O}\left(\Tilde{\alpha}^{3}\right)
\end{equation}
where
\begin{equation}
\omega_{\mathrm{GR}}=\left(V_{\mathrm{eff}}^{A}-G\right)^{1 / 2}
\end{equation}
and
{\small
\begin{equation}
\delta \omega=\pm \frac{1 }{2\left[V_{\text {eff }}^{A}-G\right]^{1 / 2}}\left[\chi\,m\,(A_2-G_A)+\frac{I H}{\left(V_{\text {eff }}^{A}-V_{\text {eff }}^{S}\right)} 
\right]
\end{equation}
}
It is important to note that the above Fermi analysis result is different from that of Ref.~\cite{Wagle:2021tam}. In Ref.~\cite{Wagle:2021tam}, the authors did not include the $(A_2-G_A)$ term. 
Thus, even using Fermi estimate analysis, their analysis ignores certain terms that contribute to the correction in $\omega$ at $O(\Tilde{\alpha}^2)$.
%
%
%
%
%
\bibliography{apssamp}

\begin{thebibliography}{62}%
\makeatletter
\providecommand \@ifxundefined [1]{%
 \@ifx{#1\undefined}
}%
\providecommand \@ifnum [1]{%
 \ifnum #1\expandafter \@firstoftwo
 \else \expandafter \@secondoftwo
 \fi
}%
\providecommand \@ifx [1]{%
 \ifx #1\expandafter \@firstoftwo
 \else \expandafter \@secondoftwo
 \fi
}%
\providecommand \natexlab [1]{#1}%
\providecommand \enquote  [1]{``#1''}%
\providecommand \bibnamefont  [1]{#1}%
\providecommand \bibfnamefont [1]{#1}%
\providecommand \citenamefont [1]{#1}%
\providecommand \href@noop [0]{\@secondoftwo}%
\providecommand \href [0]{\begingroup \@sanitize@url \@href}%
\providecommand \@href[1]{\@@startlink{#1}\@@href}%
\providecommand \@@href[1]{\endgroup#1\@@endlink}%
\providecommand \@sanitize@url [0]{\catcode `\\12\catcode `\$12\catcode
  `\&12\catcode `\#12\catcode `\^12\catcode `\_12\catcode `\%12\relax}%
\providecommand \@@startlink[1]{}%
\providecommand \@@endlink[0]{}%
\providecommand \url  [0]{\begingroup\@sanitize@url \@url }%
\providecommand \@url [1]{\endgroup\@href {#1}{\urlprefix }}%
\providecommand \urlprefix  [0]{URL }%
\providecommand \Eprint [0]{\href }%
\providecommand \doibase [0]{http://dx.doi.org/}%
\providecommand \selectlanguage [0]{\@gobble}%
\providecommand \bibinfo  [0]{\@secondoftwo}%
\providecommand \bibfield  [0]{\@secondoftwo}%
\providecommand \translation [1]{[#1]}%
\providecommand \BibitemOpen [0]{}%
\providecommand \bibitemStop [0]{}%
\providecommand \bibitemNoStop [0]{.\EOS\space}%
\providecommand \EOS [0]{\spacefactor3000\relax}%
\providecommand \BibitemShut  [1]{\csname bibitem#1\endcsname}%
\let\auto@bib@innerbib\@empty
\bibitem [{\citenamefont {Regge}\ and\ \citenamefont
  {Wheeler}(1957)}]{Regge:1957td}%
  \BibitemOpen
  \bibfield  {author} {\bibinfo {author} {\bibfnamefont {T.}~\bibnamefont
  {Regge}}\ and\ \bibinfo {author} {\bibfnamefont {J.~A.}\ \bibnamefont
  {Wheeler}},\ }\href {\doibase 10.1103/PhysRev.108.1063} {\bibfield  {journal}
  {\bibinfo  {journal} {Phys. Rev.}\ }\textbf {\bibinfo {volume} {108}},\
  \bibinfo {pages} {1063} (\bibinfo {year} {1957})}\BibitemShut {NoStop}%
\bibitem [{\citenamefont {Zerilli}(1970)}]{Zerilli}%
  \BibitemOpen
  \bibfield  {author} {\bibinfo {author} {\bibfnamefont {F.~J.}\ \bibnamefont
  {Zerilli}},\ }\href {\doibase 10.1103/PhysRevD.2.2141} {\bibfield  {journal}
  {\bibinfo  {journal} {Phys. Rev. D}\ }\textbf {\bibinfo {volume} {2}},\
  \bibinfo {pages} {2141} (\bibinfo {year} {1970})}\BibitemShut {NoStop}%
\bibitem [{\citenamefont {Vishveshwara}(1970)}]{Vishveshwara:1970zz}%
  \BibitemOpen
  \bibfield  {author} {\bibinfo {author} {\bibfnamefont {C.~V.}\ \bibnamefont
  {Vishveshwara}},\ }\href {\doibase 10.1038/227936a0} {\bibfield  {journal}
  {\bibinfo  {journal} {Nature}\ }\textbf {\bibinfo {volume} {227}},\ \bibinfo
  {pages} {936} (\bibinfo {year} {1970})}\BibitemShut {NoStop}%
\bibitem [{\citenamefont {Chandrasekhar}(1985)}]{Chandrasekhar:1985kt}%
  \BibitemOpen
  \bibfield  {author} {\bibinfo {author} {\bibfnamefont {S.}~\bibnamefont
  {Chandrasekhar}},\ }\href@noop {} {\emph {\bibinfo {title} {{The mathematical
  theory of black holes}}}}\ (\bibinfo {year} {1985})\BibitemShut {NoStop}%
\bibitem [{\citenamefont {Nollert}(1999)}]{Nollert:1999ji}%
  \BibitemOpen
  \bibfield  {author} {\bibinfo {author} {\bibfnamefont {H.-P.}\ \bibnamefont
  {Nollert}},\ }\href {\doibase 10.1088/0264-9381/16/12/201} {\bibfield
  {journal} {\bibinfo  {journal} {Class. Quant. Grav.}\ }\textbf {\bibinfo
  {volume} {16}},\ \bibinfo {pages} {R159} (\bibinfo {year}
  {1999})}\BibitemShut {NoStop}%
\bibitem [{\citenamefont {Kokkotas}\ and\ \citenamefont
  {Schmidt}(1999)}]{Kokkotas:1999bd}%
  \BibitemOpen
  \bibfield  {author} {\bibinfo {author} {\bibfnamefont {K.~D.}\ \bibnamefont
  {Kokkotas}}\ and\ \bibinfo {author} {\bibfnamefont {B.~G.}\ \bibnamefont
  {Schmidt}},\ }\href {\doibase 10.12942/lrr-1999-2} {\bibfield  {journal}
  {\bibinfo  {journal} {Living Rev. Rel.}\ }\textbf {\bibinfo {volume} {2}},\
  \bibinfo {pages} {2} (\bibinfo {year} {1999})},\ \Eprint
  {http://arxiv.org/abs/gr-qc/9909058} {arXiv:gr-qc/9909058} \BibitemShut
  {NoStop}%
\bibitem [{\citenamefont {Konoplya}\ and\ \citenamefont
  {Zhidenko}(2011)}]{Konoplya:2011qq}%
  \BibitemOpen
  \bibfield  {author} {\bibinfo {author} {\bibfnamefont {R.~A.}\ \bibnamefont
  {Konoplya}}\ and\ \bibinfo {author} {\bibfnamefont {A.}~\bibnamefont
  {Zhidenko}},\ }\href {\doibase 10.1103/RevModPhys.83.793} {\bibfield
  {journal} {\bibinfo  {journal} {Rev. Mod. Phys.}\ }\textbf {\bibinfo {volume}
  {83}},\ \bibinfo {pages} {793} (\bibinfo {year} {2011})},\ \Eprint
  {http://arxiv.org/abs/1102.4014} {arXiv:1102.4014 [gr-qc]} \BibitemShut
  {NoStop}%
\bibitem [{\citenamefont {Misner}\ \emph {et~al.}(1973)\citenamefont {Misner},
  \citenamefont {Thorne},\ and\ \citenamefont
  {Wheeler}}]{1973-Misner.etal-Gravitation}%
  \BibitemOpen
  \bibfield  {author} {\bibinfo {author} {\bibfnamefont {C.}~\bibnamefont
  {Misner}}, \bibinfo {author} {\bibfnamefont {K.}~\bibnamefont {Thorne}}, \
  and\ \bibinfo {author} {\bibfnamefont {J.}~\bibnamefont {Wheeler}},\
  }\href@noop {} {\emph {\bibinfo {title} {Gravitation}}}\ (\bibinfo
  {publisher} {Freeman},\ \bibinfo {year} {1973})\BibitemShut {NoStop}%
\bibitem [{\citenamefont {Barack}\ \emph {et~al.}(2019)\citenamefont {Barack}
  \emph {et~al.}}]{Barack:2018yly}%
  \BibitemOpen
  \bibfield  {author} {\bibinfo {author} {\bibfnamefont {L.}~\bibnamefont
  {Barack}} \emph {et~al.},\ }\href {\doibase 10.1088/1361-6382/ab0587}
  {\bibfield  {journal} {\bibinfo  {journal} {Class. Quant. Grav.}\ }\textbf
  {\bibinfo {volume} {36}},\ \bibinfo {pages} {143001} (\bibinfo {year}
  {2019})},\ \Eprint {http://arxiv.org/abs/1806.05195} {arXiv:1806.05195
  [gr-qc]} \BibitemShut {NoStop}%
\bibitem [{\citenamefont {Abbott}\ \emph
  {et~al.}(2020{\natexlab{a}})\citenamefont {Abbott} \emph
  {et~al.}}]{Abbott:2020niy}%
  \BibitemOpen
  \bibfield  {author} {\bibinfo {author} {\bibfnamefont {R.}~\bibnamefont
  {Abbott}} \emph {et~al.} (\bibinfo {collaboration} {LIGO Scientific,
  Virgo}),\ }\href@noop {} {\  (\bibinfo {year} {2020}{\natexlab{a}})},\
  \Eprint {http://arxiv.org/abs/2010.14527} {arXiv:2010.14527 [gr-qc]}
  \BibitemShut {NoStop}%
\bibitem [{\citenamefont {Bhagwat}\ \emph {et~al.}(2020)\citenamefont
  {Bhagwat}, \citenamefont {Cabero}, \citenamefont {Capano}, \citenamefont
  {Krishnan},\ and\ \citenamefont {Brown}}]{Bhagwat:2019bwv}%
  \BibitemOpen
  \bibfield  {author} {\bibinfo {author} {\bibfnamefont {S.}~\bibnamefont
  {Bhagwat}}, \bibinfo {author} {\bibfnamefont {M.}~\bibnamefont {Cabero}},
  \bibinfo {author} {\bibfnamefont {C.~D.}\ \bibnamefont {Capano}}, \bibinfo
  {author} {\bibfnamefont {B.}~\bibnamefont {Krishnan}}, \ and\ \bibinfo
  {author} {\bibfnamefont {D.~A.}\ \bibnamefont {Brown}},\ }\href {\doibase
  10.1103/PhysRevD.102.024023} {\bibfield  {journal} {\bibinfo  {journal}
  {Phys. Rev. D}\ }\textbf {\bibinfo {volume} {102}},\ \bibinfo {pages}
  {024023} (\bibinfo {year} {2020})},\ \Eprint
  {http://arxiv.org/abs/1910.13203} {arXiv:1910.13203 [gr-qc]} \BibitemShut
  {NoStop}%
\bibitem [{\citenamefont {Venumadhav}\ \emph {et~al.}(2020)\citenamefont
  {Venumadhav}, \citenamefont {Zackay}, \citenamefont {Roulet}, \citenamefont
  {Dai},\ and\ \citenamefont {Zaldarriaga}}]{Venumadhav:2019lyq}%
  \BibitemOpen
  \bibfield  {author} {\bibinfo {author} {\bibfnamefont {T.}~\bibnamefont
  {Venumadhav}}, \bibinfo {author} {\bibfnamefont {B.}~\bibnamefont {Zackay}},
  \bibinfo {author} {\bibfnamefont {J.}~\bibnamefont {Roulet}}, \bibinfo
  {author} {\bibfnamefont {L.}~\bibnamefont {Dai}}, \ and\ \bibinfo {author}
  {\bibfnamefont {M.}~\bibnamefont {Zaldarriaga}},\ }\href {\doibase
  10.1103/PhysRevD.101.083030} {\bibfield  {journal} {\bibinfo  {journal}
  {Phys. Rev. D}\ }\textbf {\bibinfo {volume} {101}},\ \bibinfo {pages}
  {083030} (\bibinfo {year} {2020})},\ \Eprint
  {http://arxiv.org/abs/1904.07214} {arXiv:1904.07214 [astro-ph.HE]}
  \BibitemShut {NoStop}%
\bibitem [{\citenamefont {Abbott}\ \emph
  {et~al.}(2020{\natexlab{b}})\citenamefont {Abbott} \emph {et~al.}}]{LIGOSNR}%
  \BibitemOpen
  \bibfield  {author} {\bibinfo {author} {\bibfnamefont {R.}~\bibnamefont
  {Abbott}} \emph {et~al.} (\bibinfo {collaboration} {LIGO Scientific,
  Virgo}),\ }\href@noop {} {\  (\bibinfo {year} {2020}{\natexlab{b}})},\
  \Eprint {http://arxiv.org/abs/2010.14529} {arXiv:2010.14529 [gr-qc]}
  \BibitemShut {NoStop}%
\bibitem [{\citenamefont {Isi}\ \emph {et~al.}(2019)\citenamefont {Isi},
  \citenamefont {Giesler}, \citenamefont {Farr}, \citenamefont {Scheel},\ and\
  \citenamefont {Teukolsky}}]{Isi:2019aib}%
  \BibitemOpen
  \bibfield  {author} {\bibinfo {author} {\bibfnamefont {M.}~\bibnamefont
  {Isi}}, \bibinfo {author} {\bibfnamefont {M.}~\bibnamefont {Giesler}},
  \bibinfo {author} {\bibfnamefont {W.~M.}\ \bibnamefont {Farr}}, \bibinfo
  {author} {\bibfnamefont {M.~A.}\ \bibnamefont {Scheel}}, \ and\ \bibinfo
  {author} {\bibfnamefont {S.~A.}\ \bibnamefont {Teukolsky}},\ }\href {\doibase
  10.1103/PhysRevLett.123.111102} {\bibfield  {journal} {\bibinfo  {journal}
  {Phys. Rev. Lett.}\ }\textbf {\bibinfo {volume} {123}},\ \bibinfo {pages}
  {111102} (\bibinfo {year} {2019})},\ \Eprint
  {http://arxiv.org/abs/1905.00869} {arXiv:1905.00869 [gr-qc]} \BibitemShut
  {NoStop}%
\bibitem [{\citenamefont {Dhanpal}\ \emph {et~al.}(2019)\citenamefont
  {Dhanpal}, \citenamefont {Ghosh}, \citenamefont {Mehta}, \citenamefont
  {Ajith},\ and\ \citenamefont {Sathyaprakash}}]{Dhanpal:2018ufk}%
  \BibitemOpen
  \bibfield  {author} {\bibinfo {author} {\bibfnamefont {S.}~\bibnamefont
  {Dhanpal}}, \bibinfo {author} {\bibfnamefont {A.}~\bibnamefont {Ghosh}},
  \bibinfo {author} {\bibfnamefont {A.~K.}\ \bibnamefont {Mehta}}, \bibinfo
  {author} {\bibfnamefont {P.}~\bibnamefont {Ajith}}, \ and\ \bibinfo {author}
  {\bibfnamefont {B.~S.}\ \bibnamefont {Sathyaprakash}},\ }\href {\doibase
  10.1103/PhysRevD.99.104056} {\bibfield  {journal} {\bibinfo  {journal} {Phys.
  Rev. D}\ }\textbf {\bibinfo {volume} {99}},\ \bibinfo {pages} {104056}
  (\bibinfo {year} {2019})},\ \Eprint {http://arxiv.org/abs/1804.03297}
  {arXiv:1804.03297 [gr-qc]} \BibitemShut {NoStop}%
\bibitem [{\citenamefont {Baibhav}\ \emph {et~al.}(2018)\citenamefont
  {Baibhav}, \citenamefont {Berti}, \citenamefont {Cardoso},\ and\
  \citenamefont {Khanna}}]{Cardoso_Request_Overtone_PhysRevD.97.044048}%
  \BibitemOpen
  \bibfield  {author} {\bibinfo {author} {\bibfnamefont {V.}~\bibnamefont
  {Baibhav}}, \bibinfo {author} {\bibfnamefont {E.}~\bibnamefont {Berti}},
  \bibinfo {author} {\bibfnamefont {V.}~\bibnamefont {Cardoso}}, \ and\
  \bibinfo {author} {\bibfnamefont {G.}~\bibnamefont {Khanna}},\ }\href
  {\doibase 10.1103/PhysRevD.97.044048} {\bibfield  {journal} {\bibinfo
  {journal} {Phys. Rev. D}\ }\textbf {\bibinfo {volume} {97}},\ \bibinfo
  {pages} {044048} (\bibinfo {year} {2018})}\BibitemShut {NoStop}%
\bibitem [{\citenamefont {Abbott}\ \emph
  {et~al.}(2017{\natexlab{a}})\citenamefont {Abbott} \emph
  {et~al.}}]{Evans:2016mbw}%
  \BibitemOpen
  \bibfield  {author} {\bibinfo {author} {\bibfnamefont {B.~P.}\ \bibnamefont
  {Abbott}} \emph {et~al.} (\bibinfo {collaboration} {LIGO Scientific}),\
  }\href {\doibase 10.1088/1361-6382/aa51f4} {\bibfield  {journal} {\bibinfo
  {journal} {Class. Quant. Grav.}\ }\textbf {\bibinfo {volume} {34}},\ \bibinfo
  {pages} {044001} (\bibinfo {year} {2017}{\natexlab{a}})},\ \Eprint
  {http://arxiv.org/abs/1607.08697} {arXiv:1607.08697 [astro-ph.IM]}
  \BibitemShut {NoStop}%
\bibitem [{\citenamefont {Berti}\ \emph {et~al.}(2016)\citenamefont {Berti},
  \citenamefont {Sesana}, \citenamefont {Barausse}, \citenamefont {Cardoso},\
  and\ \citenamefont {Belczynski}}]{Cardoso_Request_1_PhysRevLett.117.101102}%
  \BibitemOpen
  \bibfield  {author} {\bibinfo {author} {\bibfnamefont {E.}~\bibnamefont
  {Berti}}, \bibinfo {author} {\bibfnamefont {A.}~\bibnamefont {Sesana}},
  \bibinfo {author} {\bibfnamefont {E.}~\bibnamefont {Barausse}}, \bibinfo
  {author} {\bibfnamefont {V.}~\bibnamefont {Cardoso}}, \ and\ \bibinfo
  {author} {\bibfnamefont {K.}~\bibnamefont {Belczynski}},\ }\href {\doibase
  10.1103/PhysRevLett.117.101102} {\bibfield  {journal} {\bibinfo  {journal}
  {Phys. Rev. Lett.}\ }\textbf {\bibinfo {volume} {117}},\ \bibinfo {pages}
  {101102} (\bibinfo {year} {2016})}\BibitemShut {NoStop}%
\bibitem [{\citenamefont {Nakano}\ \emph {et~al.}(2015)\citenamefont {Nakano},
  \citenamefont {Tanaka},\ and\ \citenamefont {Nakamura}}]{Nakano:2015uja}%
  \BibitemOpen
  \bibfield  {author} {\bibinfo {author} {\bibfnamefont {H.}~\bibnamefont
  {Nakano}}, \bibinfo {author} {\bibfnamefont {T.}~\bibnamefont {Tanaka}}, \
  and\ \bibinfo {author} {\bibfnamefont {T.}~\bibnamefont {Nakamura}},\ }\href
  {\doibase 10.1103/PhysRevD.92.064003} {\bibfield  {journal} {\bibinfo
  {journal} {Phys. Rev. D}\ }\textbf {\bibinfo {volume} {92}},\ \bibinfo
  {pages} {064003} (\bibinfo {year} {2015})},\ \Eprint
  {http://arxiv.org/abs/1506.00560} {arXiv:1506.00560 [astro-ph.HE]}
  \BibitemShut {NoStop}%
\bibitem [{\citenamefont {Maselli}\ \emph {et~al.}(2020)\citenamefont
  {Maselli}, \citenamefont {Pani}, \citenamefont {Gualtieri},\ and\
  \citenamefont {Berti}}]{Berti_PhysRevD.101.024043}%
  \BibitemOpen
  \bibfield  {author} {\bibinfo {author} {\bibfnamefont {A.}~\bibnamefont
  {Maselli}}, \bibinfo {author} {\bibfnamefont {P.}~\bibnamefont {Pani}},
  \bibinfo {author} {\bibfnamefont {L.}~\bibnamefont {Gualtieri}}, \ and\
  \bibinfo {author} {\bibfnamefont {E.}~\bibnamefont {Berti}},\ }\href
  {\doibase 10.1103/PhysRevD.101.024043} {\bibfield  {journal} {\bibinfo
  {journal} {Phys. Rev. D}\ }\textbf {\bibinfo {volume} {101}},\ \bibinfo
  {pages} {024043} (\bibinfo {year} {2020})}\BibitemShut {NoStop}%
\bibitem [{\citenamefont {McManus}\ \emph {et~al.}(2019)\citenamefont
  {McManus}, \citenamefont {Berti}, \citenamefont {Macedo}, \citenamefont
  {Kimura}, \citenamefont {Maselli},\ and\ \citenamefont
  {Cardoso}}]{Berti_2_PhysRevD.100.044061}%
  \BibitemOpen
  \bibfield  {author} {\bibinfo {author} {\bibfnamefont {R.}~\bibnamefont
  {McManus}}, \bibinfo {author} {\bibfnamefont {E.}~\bibnamefont {Berti}},
  \bibinfo {author} {\bibfnamefont {C.~F.~B.}\ \bibnamefont {Macedo}}, \bibinfo
  {author} {\bibfnamefont {M.}~\bibnamefont {Kimura}}, \bibinfo {author}
  {\bibfnamefont {A.}~\bibnamefont {Maselli}}, \ and\ \bibinfo {author}
  {\bibfnamefont {V.}~\bibnamefont {Cardoso}},\ }\href {\doibase
  10.1103/PhysRevD.100.044061} {\bibfield  {journal} {\bibinfo  {journal}
  {Phys. Rev. D}\ }\textbf {\bibinfo {volume} {100}},\ \bibinfo {pages}
  {044061} (\bibinfo {year} {2019})}\BibitemShut {NoStop}%
\bibitem [{\citenamefont {Stelle}(1977)}]{Stelle:1976gc}%
  \BibitemOpen
  \bibfield  {author} {\bibinfo {author} {\bibfnamefont {K.~S.}\ \bibnamefont
  {Stelle}},\ }\href {\doibase 10.1103/PhysRevD.16.953} {\bibfield  {journal}
  {\bibinfo  {journal} {Phys. Rev. D}\ }\textbf {\bibinfo {volume} {16}},\
  \bibinfo {pages} {953} (\bibinfo {year} {1977})}\BibitemShut {NoStop}%
\bibitem [{\citenamefont {De~Felice}\ and\ \citenamefont
  {Tsujikawa}(2010)}]{DeFelice:2010aj}%
  \BibitemOpen
  \bibfield  {author} {\bibinfo {author} {\bibfnamefont {A.}~\bibnamefont
  {De~Felice}}\ and\ \bibinfo {author} {\bibfnamefont {S.}~\bibnamefont
  {Tsujikawa}},\ }\href {\doibase 10.12942/lrr-2010-3} {\bibfield  {journal}
  {\bibinfo  {journal} {Living Rev. Rel.}\ }\textbf {\bibinfo {volume} {13}},\
  \bibinfo {pages} {3} (\bibinfo {year} {2010})},\ \Eprint
  {http://arxiv.org/abs/1002.4928} {arXiv:1002.4928 [gr-qc]} \BibitemShut
  {NoStop}%
\bibitem [{\citenamefont {Sotiriou}\ and\ \citenamefont
  {Faraoni}(2010)}]{Sotiriou:2008rp}%
  \BibitemOpen
  \bibfield  {author} {\bibinfo {author} {\bibfnamefont {T.~P.}\ \bibnamefont
  {Sotiriou}}\ and\ \bibinfo {author} {\bibfnamefont {V.}~\bibnamefont
  {Faraoni}},\ }\href {\doibase 10.1103/RevModPhys.82.451} {\bibfield
  {journal} {\bibinfo  {journal} {Rev. Mod. Phys.}\ }\textbf {\bibinfo {volume}
  {82}},\ \bibinfo {pages} {451} (\bibinfo {year} {2010})},\ \Eprint
  {http://arxiv.org/abs/0805.1726} {arXiv:0805.1726 [gr-qc]} \BibitemShut
  {NoStop}%
\bibitem [{\citenamefont {Capozziello}\ and\ \citenamefont
  {De~Laurentis}(2011)}]{Capozziello:2011et}%
  \BibitemOpen
  \bibfield  {author} {\bibinfo {author} {\bibfnamefont {S.}~\bibnamefont
  {Capozziello}}\ and\ \bibinfo {author} {\bibfnamefont {M.}~\bibnamefont
  {De~Laurentis}},\ }\href {\doibase 10.1016/j.physrep.2011.09.003} {\bibfield
  {journal} {\bibinfo  {journal} {Phys. Rept.}\ }\textbf {\bibinfo {volume}
  {509}},\ \bibinfo {pages} {167} (\bibinfo {year} {2011})},\ \Eprint
  {http://arxiv.org/abs/1108.6266} {arXiv:1108.6266 [gr-qc]} \BibitemShut
  {NoStop}%
\bibitem [{\citenamefont {de~Rham}(2014)}]{deRham:2014zqa}%
  \BibitemOpen
  \bibfield  {author} {\bibinfo {author} {\bibfnamefont {C.}~\bibnamefont
  {de~Rham}},\ }\href {\doibase 10.12942/lrr-2014-7} {\bibfield  {journal}
  {\bibinfo  {journal} {Living Rev. Rel.}\ }\textbf {\bibinfo {volume} {17}},\
  \bibinfo {pages} {7} (\bibinfo {year} {2014})},\ \Eprint
  {http://arxiv.org/abs/1401.4173} {arXiv:1401.4173 [hep-th]} \BibitemShut
  {NoStop}%
\bibitem [{\citenamefont {Alexander}\ and\ \citenamefont
  {Yunes}(2009)}]{Alexander:2009tp}%
  \BibitemOpen
  \bibfield  {author} {\bibinfo {author} {\bibfnamefont {S.}~\bibnamefont
  {Alexander}}\ and\ \bibinfo {author} {\bibfnamefont {N.}~\bibnamefont
  {Yunes}},\ }\href {\doibase 10.1016/j.physrep.2009.07.002} {\bibfield
  {journal} {\bibinfo  {journal} {Phys. Rept.}\ }\textbf {\bibinfo {volume}
  {480}},\ \bibinfo {pages} {1} (\bibinfo {year} {2009})},\ \Eprint
  {http://arxiv.org/abs/0907.2562} {arXiv:0907.2562 [hep-th]} \BibitemShut
  {NoStop}%
\bibitem [{\citenamefont {Donoghue}(1994)}]{Donoghue:1994dn}%
  \BibitemOpen
  \bibfield  {author} {\bibinfo {author} {\bibfnamefont {J.~F.}\ \bibnamefont
  {Donoghue}},\ }\href {\doibase 10.1103/PhysRevD.50.3874} {\bibfield
  {journal} {\bibinfo  {journal} {Phys. Rev. D}\ }\textbf {\bibinfo {volume}
  {50}},\ \bibinfo {pages} {3874} (\bibinfo {year} {1994})},\ \Eprint
  {http://arxiv.org/abs/gr-qc/9405057} {arXiv:gr-qc/9405057} \BibitemShut
  {NoStop}%
\bibitem [{\citenamefont {Will}(2006)}]{Will:2005va}%
  \BibitemOpen
  \bibfield  {author} {\bibinfo {author} {\bibfnamefont {C.~M.}\ \bibnamefont
  {Will}},\ }\href {\doibase 10.12942/lrr-2006-3} {\bibfield  {journal}
  {\bibinfo  {journal} {Living Rev. Rel.}\ }\textbf {\bibinfo {volume} {9}},\
  \bibinfo {pages} {3} (\bibinfo {year} {2006})},\ \Eprint
  {http://arxiv.org/abs/gr-qc/0510072} {arXiv:gr-qc/0510072} \BibitemShut
  {NoStop}%
\bibitem [{\citenamefont {Will}(2018)}]{Will:2018}%
  \BibitemOpen
  \bibfield  {author} {\bibinfo {author} {\bibfnamefont {C.~M.}\ \bibnamefont
  {Will}},\ }\href {\doibase 10.1017/9781316338612} {\emph {\bibinfo {title}
  {Theory and Experiment in Gravitational Physics}}},\ \bibinfo {edition}
  {2nd}\ ed.\ (\bibinfo  {publisher} {Cambridge University Press},\ \bibinfo
  {year} {2018})\BibitemShut {NoStop}%
\bibitem [{\citenamefont {Jackiw}\ and\ \citenamefont
  {Pi}(2003)}]{Jackiw:2003pm}%
  \BibitemOpen
  \bibfield  {author} {\bibinfo {author} {\bibfnamefont {R.}~\bibnamefont
  {Jackiw}}\ and\ \bibinfo {author} {\bibfnamefont {S.}~\bibnamefont {Pi}},\
  }\href {\doibase 10.1103/PhysRevD.68.104012} {\bibfield  {journal} {\bibinfo
  {journal} {Phys. Rev. D}\ }\textbf {\bibinfo {volume} {68}},\ \bibinfo
  {pages} {104012} (\bibinfo {year} {2003})},\ \Eprint
  {http://arxiv.org/abs/gr-qc/0308071} {arXiv:gr-qc/0308071} \BibitemShut
  {NoStop}%
\bibitem [{\citenamefont {Smith}\ \emph {et~al.}(2008)\citenamefont {Smith},
  \citenamefont {Erickcek}, \citenamefont {Caldwell},\ and\ \citenamefont
  {Kamionkowski}}]{Smith:2007jm}%
  \BibitemOpen
  \bibfield  {author} {\bibinfo {author} {\bibfnamefont {T.~L.}\ \bibnamefont
  {Smith}}, \bibinfo {author} {\bibfnamefont {A.~L.}\ \bibnamefont {Erickcek}},
  \bibinfo {author} {\bibfnamefont {R.~R.}\ \bibnamefont {Caldwell}}, \ and\
  \bibinfo {author} {\bibfnamefont {M.}~\bibnamefont {Kamionkowski}},\ }\href
  {\doibase 10.1103/PhysRevD.77.024015} {\bibfield  {journal} {\bibinfo
  {journal} {Phys. Rev. D}\ }\textbf {\bibinfo {volume} {77}},\ \bibinfo
  {pages} {024015} (\bibinfo {year} {2008})},\ \Eprint
  {http://arxiv.org/abs/0708.0001} {arXiv:0708.0001 [astro-ph]} \BibitemShut
  {NoStop}%
\bibitem [{\citenamefont {Delsate}\ \emph {et~al.}(2018)\citenamefont
  {Delsate}, \citenamefont {Herdeiro},\ and\ \citenamefont
  {Radu}}]{Delsate:2018ome}%
  \BibitemOpen
  \bibfield  {author} {\bibinfo {author} {\bibfnamefont {T.}~\bibnamefont
  {Delsate}}, \bibinfo {author} {\bibfnamefont {C.}~\bibnamefont {Herdeiro}}, \
  and\ \bibinfo {author} {\bibfnamefont {E.}~\bibnamefont {Radu}},\ }\href
  {\doibase 10.1016/j.physletb.2018.09.060} {\bibfield  {journal} {\bibinfo
  {journal} {Phys. Lett. B}\ }\textbf {\bibinfo {volume} {787}},\ \bibinfo
  {pages} {8} (\bibinfo {year} {2018})},\ \Eprint
  {http://arxiv.org/abs/1806.06700} {arXiv:1806.06700 [gr-qc]} \BibitemShut
  {NoStop}%
\bibitem [{\citenamefont {Yunes}\ and\ \citenamefont
  {Pretorius}(2009)}]{Yunes:2009hc}%
  \BibitemOpen
  \bibfield  {author} {\bibinfo {author} {\bibfnamefont {N.}~\bibnamefont
  {Yunes}}\ and\ \bibinfo {author} {\bibfnamefont {F.}~\bibnamefont
  {Pretorius}},\ }\href {\doibase 10.1103/PhysRevD.79.084043} {\bibfield
  {journal} {\bibinfo  {journal} {Phys. Rev. D}\ }\textbf {\bibinfo {volume}
  {79}},\ \bibinfo {pages} {084043} (\bibinfo {year} {2009})},\ \Eprint
  {http://arxiv.org/abs/0902.4669} {arXiv:0902.4669 [gr-qc]} \BibitemShut
  {NoStop}%
\bibitem [{\citenamefont {Konno}\ \emph {et~al.}(2009)\citenamefont {Konno},
  \citenamefont {Matsuyama},\ and\ \citenamefont
  {Tanda}}]{Konno10.1143/PTP.122.561}%
  \BibitemOpen
  \bibfield  {author} {\bibinfo {author} {\bibfnamefont {K.}~\bibnamefont
  {Konno}}, \bibinfo {author} {\bibfnamefont {T.}~\bibnamefont {Matsuyama}}, \
  and\ \bibinfo {author} {\bibfnamefont {S.}~\bibnamefont {Tanda}},\ }\href
  {\doibase 10.1143/PTP.122.561} {\bibfield  {journal} {\bibinfo  {journal}
  {Progress of Theoretical Physics}\ }\textbf {\bibinfo {volume} {122}},\
  \bibinfo {pages} {561} (\bibinfo {year} {2009})},\ \Eprint
  {http://arxiv.org/abs/https://academic.oup.com/ptp/article-pdf/122/2/561/9681244/122-2-561.pdf}
  {https://academic.oup.com/ptp/article-pdf/122/2/561/9681244/122-2-561.pdf}
  \BibitemShut {NoStop}%
\bibitem [{\citenamefont {Yagi}\ \emph
  {et~al.}(2012{\natexlab{a}})\citenamefont {Yagi}, \citenamefont {Yunes},\
  and\ \citenamefont {Tanaka}}]{Yagi_Spin2_PhysRevD.86.044037}%
  \BibitemOpen
  \bibfield  {author} {\bibinfo {author} {\bibfnamefont {K.}~\bibnamefont
  {Yagi}}, \bibinfo {author} {\bibfnamefont {N.}~\bibnamefont {Yunes}}, \ and\
  \bibinfo {author} {\bibfnamefont {T.}~\bibnamefont {Tanaka}},\ }\href
  {\doibase 10.1103/PhysRevD.86.044037} {\bibfield  {journal} {\bibinfo
  {journal} {Phys. Rev. D}\ }\textbf {\bibinfo {volume} {86}},\ \bibinfo
  {pages} {044037} (\bibinfo {year} {2012}{\natexlab{a}})}\BibitemShut
  {NoStop}%
\bibitem [{\citenamefont {Cano}\ and\ \citenamefont
  {Ruip\'erez}(2019)}]{Cano:2019ore}%
  \BibitemOpen
  \bibfield  {author} {\bibinfo {author} {\bibfnamefont {P.~A.}\ \bibnamefont
  {Cano}}\ and\ \bibinfo {author} {\bibfnamefont {A.}~\bibnamefont
  {Ruip\'erez}},\ }\href {\doibase 10.1007/JHEP05(2019)189} {\bibfield
  {journal} {\bibinfo  {journal} {JHEP}\ }\textbf {\bibinfo {volume} {05}},\
  \bibinfo {pages} {189} (\bibinfo {year} {2019})},\ \bibinfo {note} {[Erratum:
  JHEP 03, 187 (2020)]},\ \Eprint {http://arxiv.org/abs/1901.01315}
  {arXiv:1901.01315 [gr-qc]} \BibitemShut {NoStop}%
\bibitem [{\citenamefont {Gao}\ \emph {et~al.}(2019)\citenamefont {Gao},
  \citenamefont {Huang},\ and\ \citenamefont {Liu}}]{Gao:2018acg}%
  \BibitemOpen
  \bibfield  {author} {\bibinfo {author} {\bibfnamefont {Y.-X.}\ \bibnamefont
  {Gao}}, \bibinfo {author} {\bibfnamefont {Y.}~\bibnamefont {Huang}}, \ and\
  \bibinfo {author} {\bibfnamefont {D.-J.}\ \bibnamefont {Liu}},\ }\href
  {\doibase 10.1103/PhysRevD.99.044020} {\bibfield  {journal} {\bibinfo
  {journal} {Phys. Rev. D}\ }\textbf {\bibinfo {volume} {99}},\ \bibinfo
  {pages} {044020} (\bibinfo {year} {2019})},\ \Eprint
  {http://arxiv.org/abs/1808.01433} {arXiv:1808.01433 [gr-qc]} \BibitemShut
  {NoStop}%
\bibitem [{\citenamefont {Doneva}\ and\ \citenamefont
  {Yazadjiev}(2021)}]{Doneva:2021dcc}%
  \BibitemOpen
  \bibfield  {author} {\bibinfo {author} {\bibfnamefont {D.~D.}\ \bibnamefont
  {Doneva}}\ and\ \bibinfo {author} {\bibfnamefont {S.~S.}\ \bibnamefont
  {Yazadjiev}},\ }\href {\doibase 10.1103/PhysRevD.103.083007} {\bibfield
  {journal} {\bibinfo  {journal} {Phys. Rev. D}\ }\textbf {\bibinfo {volume}
  {103}},\ \bibinfo {pages} {083007} (\bibinfo {year} {2021})},\ \Eprint
  {http://arxiv.org/abs/2102.03940} {arXiv:2102.03940 [gr-qc]} \BibitemShut
  {NoStop}%
\bibitem [{\citenamefont {Bhattacharyya}\ and\ \citenamefont
  {Shankaranarayanan}(2019)}]{Bhattacharyya:2018hsj}%
  \BibitemOpen
  \bibfield  {author} {\bibinfo {author} {\bibfnamefont {S.}~\bibnamefont
  {Bhattacharyya}}\ and\ \bibinfo {author} {\bibfnamefont {S.}~\bibnamefont
  {Shankaranarayanan}},\ }\href {\doibase 10.1103/PhysRevD.100.024022}
  {\bibfield  {journal} {\bibinfo  {journal} {Phys. Rev. D}\ }\textbf {\bibinfo
  {volume} {100}},\ \bibinfo {pages} {024022} (\bibinfo {year} {2019})},\
  \Eprint {http://arxiv.org/abs/1812.00187} {arXiv:1812.00187 [gr-qc]}
  \BibitemShut {NoStop}%
\bibitem [{\citenamefont {Okounkova}\ \emph {et~al.}(2019)\citenamefont
  {Okounkova}, \citenamefont {Stein}, \citenamefont {Scheel},\ and\
  \citenamefont {Teukolsky}}]{Okounkova:2019dfo}%
  \BibitemOpen
  \bibfield  {author} {\bibinfo {author} {\bibfnamefont {M.}~\bibnamefont
  {Okounkova}}, \bibinfo {author} {\bibfnamefont {L.~C.}\ \bibnamefont
  {Stein}}, \bibinfo {author} {\bibfnamefont {M.~A.}\ \bibnamefont {Scheel}}, \
  and\ \bibinfo {author} {\bibfnamefont {S.~A.}\ \bibnamefont {Teukolsky}},\
  }\href {\doibase 10.1103/PhysRevD.100.104026} {\bibfield  {journal} {\bibinfo
   {journal} {Phys. Rev. D}\ }\textbf {\bibinfo {volume} {100}},\ \bibinfo
  {pages} {104026} (\bibinfo {year} {2019})},\ \Eprint
  {http://arxiv.org/abs/1906.08789} {arXiv:1906.08789 [gr-qc]} \BibitemShut
  {NoStop}%
\bibitem [{\citenamefont {Okounkova}\ \emph {et~al.}(2020)\citenamefont
  {Okounkova}, \citenamefont {Stein}, \citenamefont {Moxon}, \citenamefont
  {Scheel},\ and\ \citenamefont {Teukolsky}}]{Okounkova:2019zjf}%
  \BibitemOpen
  \bibfield  {author} {\bibinfo {author} {\bibfnamefont {M.}~\bibnamefont
  {Okounkova}}, \bibinfo {author} {\bibfnamefont {L.~C.}\ \bibnamefont
  {Stein}}, \bibinfo {author} {\bibfnamefont {J.}~\bibnamefont {Moxon}},
  \bibinfo {author} {\bibfnamefont {M.~A.}\ \bibnamefont {Scheel}}, \ and\
  \bibinfo {author} {\bibfnamefont {S.~A.}\ \bibnamefont {Teukolsky}},\ }\href
  {\doibase 10.1103/PhysRevD.101.104016} {\bibfield  {journal} {\bibinfo
  {journal} {Phys. Rev. D}\ }\textbf {\bibinfo {volume} {101}},\ \bibinfo
  {pages} {104016} (\bibinfo {year} {2020})},\ \Eprint
  {http://arxiv.org/abs/1911.02588} {arXiv:1911.02588 [gr-qc]} \BibitemShut
  {NoStop}%
\bibitem [{\citenamefont {Wagle}\ \emph {et~al.}(2021)\citenamefont {Wagle},
  \citenamefont {Yunes},\ and\ \citenamefont {Silva}}]{Wagle:2021tam}%
  \BibitemOpen
  \bibfield  {author} {\bibinfo {author} {\bibfnamefont {P.~K.}\ \bibnamefont
  {Wagle}}, \bibinfo {author} {\bibfnamefont {N.}~\bibnamefont {Yunes}}, \ and\
  \bibinfo {author} {\bibfnamefont {H.~O.}\ \bibnamefont {Silva}},\ }\href@noop
  {} {\  (\bibinfo {year} {2021})},\ \Eprint {http://arxiv.org/abs/2103.09913}
  {arXiv:2103.09913 [gr-qc]} \BibitemShut {NoStop}%
\bibitem [{\citenamefont {Cano}\ \emph {et~al.}(2020)\citenamefont {Cano},
  \citenamefont {Fransen},\ and\ \citenamefont {Hertog}}]{Cano:2020cao}%
  \BibitemOpen
  \bibfield  {author} {\bibinfo {author} {\bibfnamefont {P.~A.}\ \bibnamefont
  {Cano}}, \bibinfo {author} {\bibfnamefont {K.}~\bibnamefont {Fransen}}, \
  and\ \bibinfo {author} {\bibfnamefont {T.}~\bibnamefont {Hertog}},\ }\href
  {\doibase 10.1103/PhysRevD.102.044047} {\bibfield  {journal} {\bibinfo
  {journal} {Phys. Rev. D}\ }\textbf {\bibinfo {volume} {102}},\ \bibinfo
  {pages} {044047} (\bibinfo {year} {2020})},\ \Eprint
  {http://arxiv.org/abs/2005.03671} {arXiv:2005.03671 [gr-qc]} \BibitemShut
  {NoStop}%
\bibitem [{\citenamefont
  {Shankaranarayanan}(2019)}]{Shankaranarayanan:2019yjx}%
  \BibitemOpen
  \bibfield  {author} {\bibinfo {author} {\bibfnamefont {S.}~\bibnamefont
  {Shankaranarayanan}},\ }\href {\doibase 10.1142/S0218271819440206} {\bibfield
   {journal} {\bibinfo  {journal} {Int. J. Mod. Phys. D}\ }\textbf {\bibinfo
  {volume} {28}},\ \bibinfo {pages} {1944020} (\bibinfo {year} {2019})},\
  \Eprint {http://arxiv.org/abs/1905.03943} {arXiv:1905.03943 [gr-qc]}
  \BibitemShut {NoStop}%
\bibitem [{\citenamefont {Kojima}(1992)}]{Kojima-PhysRevD.46.4289}%
  \BibitemOpen
  \bibfield  {author} {\bibinfo {author} {\bibfnamefont {Y.}~\bibnamefont
  {Kojima}},\ }\href {\doibase 10.1103/PhysRevD.46.4289} {\bibfield  {journal}
  {\bibinfo  {journal} {Phys. Rev. D}\ }\textbf {\bibinfo {volume} {46}},\
  \bibinfo {pages} {4289} (\bibinfo {year} {1992})}\BibitemShut {NoStop}%
\bibitem [{\citenamefont {Pani}(2013)}]{Pani:2013pma}%
  \BibitemOpen
  \bibfield  {author} {\bibinfo {author} {\bibfnamefont {P.}~\bibnamefont
  {Pani}},\ }\href {\doibase 10.1142/S0217751X13400186} {\bibfield  {journal}
  {\bibinfo  {journal} {Int. J. Mod. Phys. A}\ }\textbf {\bibinfo {volume}
  {28}},\ \bibinfo {pages} {1340018} (\bibinfo {year} {2013})},\ \Eprint
  {http://arxiv.org/abs/1305.6759} {arXiv:1305.6759 [gr-qc]} \BibitemShut
  {NoStop}%
\bibitem [{\citenamefont {Alexander}\ \emph {et~al.}(2006)\citenamefont
  {Alexander}, \citenamefont {Peskin},\ and\ \citenamefont
  {Sheikh-Jabbari}}]{Alexander:2004us}%
  \BibitemOpen
  \bibfield  {author} {\bibinfo {author} {\bibfnamefont {S.~H.-S.}\
  \bibnamefont {Alexander}}, \bibinfo {author} {\bibfnamefont {M.~E.}\
  \bibnamefont {Peskin}}, \ and\ \bibinfo {author} {\bibfnamefont {M.~M.}\
  \bibnamefont {Sheikh-Jabbari}},\ }\href {\doibase
  10.1103/PhysRevLett.96.081301} {\bibfield  {journal} {\bibinfo  {journal}
  {Phys. Rev. Lett.}\ }\textbf {\bibinfo {volume} {96}},\ \bibinfo {pages}
  {081301} (\bibinfo {year} {2006})},\ \Eprint
  {http://arxiv.org/abs/hep-th/0403069} {arXiv:hep-th/0403069} \BibitemShut
  {NoStop}%
\bibitem [{\citenamefont {Leaver}(1985)}]{Leaver:1985ax}%
  \BibitemOpen
  \bibfield  {author} {\bibinfo {author} {\bibfnamefont {E.~W.}\ \bibnamefont
  {Leaver}},\ }\href {\doibase 10.1098/rspa.1985.0119} {\bibfield  {journal}
  {\bibinfo  {journal} {Proc. Roy. Soc. Lond. A}\ }\textbf {\bibinfo {volume}
  {402}},\ \bibinfo {pages} {285} (\bibinfo {year} {1985})}\BibitemShut
  {NoStop}%
\bibitem [{\citenamefont {Mark}\ \emph {et~al.}(2015)\citenamefont {Mark},
  \citenamefont {Yang}, \citenamefont {Zimmerman},\ and\ \citenamefont
  {Chen}}]{Mark:2014aja}%
  \BibitemOpen
  \bibfield  {author} {\bibinfo {author} {\bibfnamefont {Z.}~\bibnamefont
  {Mark}}, \bibinfo {author} {\bibfnamefont {H.}~\bibnamefont {Yang}}, \bibinfo
  {author} {\bibfnamefont {A.}~\bibnamefont {Zimmerman}}, \ and\ \bibinfo
  {author} {\bibfnamefont {Y.}~\bibnamefont {Chen}},\ }\href {\doibase
  10.1103/PhysRevD.91.044025} {\bibfield  {journal} {\bibinfo  {journal} {Phys.
  Rev. D}\ }\textbf {\bibinfo {volume} {91}},\ \bibinfo {pages} {044025}
  (\bibinfo {year} {2015})},\ \Eprint {http://arxiv.org/abs/1409.5800}
  {arXiv:1409.5800 [gr-qc]} \BibitemShut {NoStop}%
\bibitem [{\citenamefont {Chandrasekhar}(1975)}]{Chandrasekhar:1975nkd}%
  \BibitemOpen
  \bibfield  {author} {\bibinfo {author} {\bibfnamefont {S.}~\bibnamefont
  {Chandrasekhar}},\ }\href {\doibase 10.1098/rspa.1975.0066} {\bibfield
  {journal} {\bibinfo  {journal} {Proc. Roy. Soc. Lond. A}\ }\textbf {\bibinfo
  {volume} {343}},\ \bibinfo {pages} {289} (\bibinfo {year}
  {1975})}\BibitemShut {NoStop}%
\bibitem [{\citenamefont {Berti}\ \emph {et~al.}(2009)\citenamefont {Berti},
  \citenamefont {Cardoso},\ and\ \citenamefont {Starinets}}]{Berti:2009kk}%
  \BibitemOpen
  \bibfield  {author} {\bibinfo {author} {\bibfnamefont {E.}~\bibnamefont
  {Berti}}, \bibinfo {author} {\bibfnamefont {V.}~\bibnamefont {Cardoso}}, \
  and\ \bibinfo {author} {\bibfnamefont {A.~O.}\ \bibnamefont {Starinets}},\
  }\href {\doibase 10.1088/0264-9381/26/16/163001} {\bibfield  {journal}
  {\bibinfo  {journal} {Class. Quant. Grav.}\ }\textbf {\bibinfo {volume}
  {26}},\ \bibinfo {pages} {163001} (\bibinfo {year} {2009})},\ \Eprint
  {http://arxiv.org/abs/0905.2975} {arXiv:0905.2975 [gr-qc]} \BibitemShut
  {NoStop}%
\bibitem [{\citenamefont {Ali-Haimoud}\ and\ \citenamefont
  {Chen}(2011)}]{yanbei:2011fw}%
  \BibitemOpen
  \bibfield  {author} {\bibinfo {author} {\bibfnamefont {Y.}~\bibnamefont
  {Ali-Haimoud}}\ and\ \bibinfo {author} {\bibfnamefont {Y.}~\bibnamefont
  {Chen}},\ }\href {\doibase 10.1103/PhysRevD.84.124033} {\bibfield  {journal}
  {\bibinfo  {journal} {Phys. Rev. D}\ }\textbf {\bibinfo {volume} {84}},\
  \bibinfo {pages} {124033} (\bibinfo {year} {2011})},\ \Eprint
  {http://arxiv.org/abs/1110.5329} {arXiv:1110.5329 [astro-ph.HE]} \BibitemShut
  {NoStop}%
\bibitem [{\citenamefont {Yagi}\ \emph
  {et~al.}(2012{\natexlab{b}})\citenamefont {Yagi}, \citenamefont {Yunes},\
  and\ \citenamefont {Tanaka}}]{Yagi:2012ya}%
  \BibitemOpen
  \bibfield  {author} {\bibinfo {author} {\bibfnamefont {K.}~\bibnamefont
  {Yagi}}, \bibinfo {author} {\bibfnamefont {N.}~\bibnamefont {Yunes}}, \ and\
  \bibinfo {author} {\bibfnamefont {T.}~\bibnamefont {Tanaka}},\ }\href
  {\doibase 10.1103/PhysRevD.86.044037} {\bibfield  {journal} {\bibinfo
  {journal} {Phys. Rev. D}\ }\textbf {\bibinfo {volume} {86}},\ \bibinfo
  {pages} {044037} (\bibinfo {year} {2012}{\natexlab{b}})},\ \bibinfo {note}
  {[Erratum: Phys.Rev.D 89, 049902 (2014)]},\ \Eprint
  {http://arxiv.org/abs/1206.6130} {arXiv:1206.6130 [gr-qc]} \BibitemShut
  {NoStop}%
\bibitem [{\citenamefont {Riley}\ \emph {et~al.}(2019)\citenamefont {Riley},
  \citenamefont {Watts}, \citenamefont {Bogdanov}, \citenamefont {Ray},
  \citenamefont {Ludlam}, \citenamefont {Guillot}, \citenamefont {Arzoumanian},
  \citenamefont {Baker}, \citenamefont {Bilous}, \citenamefont {Chakrabarty},
  \citenamefont {Gendreau}, \citenamefont {Harding}, \citenamefont {Ho},
  \citenamefont {Lattimer}, \citenamefont {Morsink},\ and\ \citenamefont
  {Strohmayer}}]{Riley_2019}%
  \BibitemOpen
  \bibfield  {author} {\bibinfo {author} {\bibfnamefont {T.~E.}\ \bibnamefont
  {Riley}}, \bibinfo {author} {\bibfnamefont {A.~L.}\ \bibnamefont {Watts}},
  \bibinfo {author} {\bibfnamefont {S.}~\bibnamefont {Bogdanov}}, \bibinfo
  {author} {\bibfnamefont {P.~S.}\ \bibnamefont {Ray}}, \bibinfo {author}
  {\bibfnamefont {R.~M.}\ \bibnamefont {Ludlam}}, \bibinfo {author}
  {\bibfnamefont {S.}~\bibnamefont {Guillot}}, \bibinfo {author} {\bibfnamefont
  {Z.}~\bibnamefont {Arzoumanian}}, \bibinfo {author} {\bibfnamefont {C.~L.}\
  \bibnamefont {Baker}}, \bibinfo {author} {\bibfnamefont {A.~V.}\ \bibnamefont
  {Bilous}}, \bibinfo {author} {\bibfnamefont {D.}~\bibnamefont {Chakrabarty}},
  \bibinfo {author} {\bibfnamefont {K.~C.}\ \bibnamefont {Gendreau}}, \bibinfo
  {author} {\bibfnamefont {A.~K.}\ \bibnamefont {Harding}}, \bibinfo {author}
  {\bibfnamefont {W.~C.~G.}\ \bibnamefont {Ho}}, \bibinfo {author}
  {\bibfnamefont {J.~M.}\ \bibnamefont {Lattimer}}, \bibinfo {author}
  {\bibfnamefont {S.~M.}\ \bibnamefont {Morsink}}, \ and\ \bibinfo {author}
  {\bibfnamefont {T.~E.}\ \bibnamefont {Strohmayer}},\ }\href {\doibase
  10.3847/2041-8213/ab481c} {\bibfield  {journal} {\bibinfo  {journal} {The
  Astrophysical Journal}\ }\textbf {\bibinfo {volume} {887}},\ \bibinfo {pages}
  {L21} (\bibinfo {year} {2019})}\BibitemShut {NoStop}%
\bibitem [{\citenamefont {Miller}\ \emph {et~al.}(2019)\citenamefont {Miller},
  \citenamefont {Lamb}, \citenamefont {Dittmann}, \citenamefont {Bogdanov},
  \citenamefont {Arzoumanian}, \citenamefont {Gendreau}, \citenamefont
  {Guillot}, \citenamefont {Harding}, \citenamefont {Ho}, \citenamefont
  {Lattimer}, \citenamefont {Ludlam}, \citenamefont {Mahmoodifar},
  \citenamefont {Morsink}, \citenamefont {Ray}, \citenamefont {Strohmayer},
  \citenamefont {Wood}, \citenamefont {Enoto}, \citenamefont {Foster},
  \citenamefont {Okajima}, \citenamefont {Prigozhin},\ and\ \citenamefont
  {Soong}}]{Miller_2019}%
  \BibitemOpen
  \bibfield  {author} {\bibinfo {author} {\bibfnamefont {M.~C.}\ \bibnamefont
  {Miller}}, \bibinfo {author} {\bibfnamefont {F.~K.}\ \bibnamefont {Lamb}},
  \bibinfo {author} {\bibfnamefont {A.~J.}\ \bibnamefont {Dittmann}}, \bibinfo
  {author} {\bibfnamefont {S.}~\bibnamefont {Bogdanov}}, \bibinfo {author}
  {\bibfnamefont {Z.}~\bibnamefont {Arzoumanian}}, \bibinfo {author}
  {\bibfnamefont {K.~C.}\ \bibnamefont {Gendreau}}, \bibinfo {author}
  {\bibfnamefont {S.}~\bibnamefont {Guillot}}, \bibinfo {author} {\bibfnamefont
  {A.~K.}\ \bibnamefont {Harding}}, \bibinfo {author} {\bibfnamefont
  {W.~C.~G.}\ \bibnamefont {Ho}}, \bibinfo {author} {\bibfnamefont {J.~M.}\
  \bibnamefont {Lattimer}}, \bibinfo {author} {\bibfnamefont {R.~M.}\
  \bibnamefont {Ludlam}}, \bibinfo {author} {\bibfnamefont {S.}~\bibnamefont
  {Mahmoodifar}}, \bibinfo {author} {\bibfnamefont {S.~M.}\ \bibnamefont
  {Morsink}}, \bibinfo {author} {\bibfnamefont {P.~S.}\ \bibnamefont {Ray}},
  \bibinfo {author} {\bibfnamefont {T.~E.}\ \bibnamefont {Strohmayer}},
  \bibinfo {author} {\bibfnamefont {K.~S.}\ \bibnamefont {Wood}}, \bibinfo
  {author} {\bibfnamefont {T.}~\bibnamefont {Enoto}}, \bibinfo {author}
  {\bibfnamefont {R.}~\bibnamefont {Foster}}, \bibinfo {author} {\bibfnamefont
  {T.}~\bibnamefont {Okajima}}, \bibinfo {author} {\bibfnamefont
  {G.}~\bibnamefont {Prigozhin}}, \ and\ \bibinfo {author} {\bibfnamefont
  {Y.}~\bibnamefont {Soong}},\ }\href {\doibase 10.3847/2041-8213/ab50c5}
  {\bibfield  {journal} {\bibinfo  {journal} {The Astrophysical Journal}\
  }\textbf {\bibinfo {volume} {887}},\ \bibinfo {pages} {L24} (\bibinfo {year}
  {2019})}\BibitemShut {NoStop}%
\bibitem [{\citenamefont {Gendreau}\ \emph {et~al.}(2016)\citenamefont
  {Gendreau} \emph {et~al.}}]{NICER10.1117/12.2231304}%
  \BibitemOpen
  \bibfield  {author} {\bibinfo {author} {\bibfnamefont {K.~C.}\ \bibnamefont
  {Gendreau}} \emph {et~al.},\ }in\ \href {\doibase 10.1117/12.2231304} {\emph
  {\bibinfo {booktitle} {Space Telescopes and Instrumentation 2016: Ultraviolet
  to Gamma Ray}}},\ Vol.\ \bibinfo {volume} {9905},\ \bibinfo {editor} {edited
  by\ \bibinfo {editor} {\bibfnamefont {J.-W.~A.}\ \bibnamefont {den Herder}},
  \bibinfo {editor} {\bibfnamefont {T.}~\bibnamefont {Takahashi}}, \ and\
  \bibinfo {editor} {\bibfnamefont {M.}~\bibnamefont {Bautz}}},\ \bibinfo
  {organization} {International Society for Optics and Photonics}\ (\bibinfo
  {publisher} {SPIE},\ \bibinfo {year} {2016})\ pp.\ \bibinfo {pages} {420 --
  435}\BibitemShut {NoStop}%
\bibitem [{\citenamefont {Silva}\ \emph {et~al.}(2021)\citenamefont {Silva},
  \citenamefont {Holgado}, \citenamefont {C\'ardenas-Avenda\~no},\ and\
  \citenamefont {Yunes}}]{Silva:2020acr}%
  \BibitemOpen
  \bibfield  {author} {\bibinfo {author} {\bibfnamefont {H.~O.}\ \bibnamefont
  {Silva}}, \bibinfo {author} {\bibfnamefont {A.~M.}\ \bibnamefont {Holgado}},
  \bibinfo {author} {\bibfnamefont {A.}~\bibnamefont {C\'ardenas-Avenda\~no}},
  \ and\ \bibinfo {author} {\bibfnamefont {N.}~\bibnamefont {Yunes}},\ }\href
  {\doibase 10.1103/PhysRevLett.126.181101} {\bibfield  {journal} {\bibinfo
  {journal} {Phys. Rev. Lett.}\ }\textbf {\bibinfo {volume} {126}},\ \bibinfo
  {pages} {181101} (\bibinfo {year} {2021})},\ \Eprint
  {http://arxiv.org/abs/2004.01253} {arXiv:2004.01253 [gr-qc]} \BibitemShut
  {NoStop}%
\bibitem [{\citenamefont {Abbott}\ \emph
  {et~al.}(2017{\natexlab{b}})\citenamefont {Abbott} \emph
  {et~al.}}]{BNSevent:2017qsa}%
  \BibitemOpen
  \bibfield  {author} {\bibinfo {author} {\bibfnamefont {B.~P.}\ \bibnamefont
  {Abbott}} \emph {et~al.} (\bibinfo {collaboration} {LIGO Scientific,
  Virgo}),\ }\href {\doibase 10.1103/PhysRevLett.119.161101} {\bibfield
  {journal} {\bibinfo  {journal} {Phys. Rev. Lett.}\ }\textbf {\bibinfo
  {volume} {119}},\ \bibinfo {pages} {161101} (\bibinfo {year}
  {2017}{\natexlab{b}})},\ \Eprint {http://arxiv.org/abs/1710.05832}
  {arXiv:1710.05832 [gr-qc]} \BibitemShut {NoStop}%
\bibitem [{\citenamefont {Bhattacharyya}\ and\ \citenamefont
  {Shankaranarayanan}(2017)}]{Bhattacharyya:2017tyc}%
  \BibitemOpen
  \bibfield  {author} {\bibinfo {author} {\bibfnamefont {S.}~\bibnamefont
  {Bhattacharyya}}\ and\ \bibinfo {author} {\bibfnamefont {S.}~\bibnamefont
  {Shankaranarayanan}},\ }\href {\doibase 10.1103/PhysRevD.96.064044}
  {\bibfield  {journal} {\bibinfo  {journal} {Phys. Rev. D}\ }\textbf {\bibinfo
  {volume} {96}},\ \bibinfo {pages} {064044} (\bibinfo {year} {2017})},\
  \Eprint {http://arxiv.org/abs/1704.07044} {arXiv:1704.07044 [gr-qc]}
  \BibitemShut {NoStop}%
\bibitem [{\citenamefont {Bhattacharyya}\ and\ \citenamefont
  {Shankaranarayanan}(2018)}]{Bhattacharyya:2018qbe}%
  \BibitemOpen
  \bibfield  {author} {\bibinfo {author} {\bibfnamefont {S.}~\bibnamefont
  {Bhattacharyya}}\ and\ \bibinfo {author} {\bibfnamefont {S.}~\bibnamefont
  {Shankaranarayanan}},\ }\href {\doibase 10.1140/epjc/s10052-018-6222-1}
  {\bibfield  {journal} {\bibinfo  {journal} {Eur. Phys. J. C}\ }\textbf
  {\bibinfo {volume} {78}},\ \bibinfo {pages} {737} (\bibinfo {year} {2018})},\
  \Eprint {http://arxiv.org/abs/1803.07576} {arXiv:1803.07576 [gr-qc]}
  \BibitemShut {NoStop}%
\bibitem [{\citenamefont {Goldberg}\ \emph {et~al.}(1967)\citenamefont
  {Goldberg}, \citenamefont {MacFarlane}, \citenamefont {Newman}, \citenamefont
  {Rohrlich},\ and\ \citenamefont {Sudarshan}}]{Goldberg:1966uu}%
  \BibitemOpen
  \bibfield  {author} {\bibinfo {author} {\bibfnamefont {J.~N.}\ \bibnamefont
  {Goldberg}}, \bibinfo {author} {\bibfnamefont {A.~J.}\ \bibnamefont
  {MacFarlane}}, \bibinfo {author} {\bibfnamefont {E.~T.}\ \bibnamefont
  {Newman}}, \bibinfo {author} {\bibfnamefont {F.}~\bibnamefont {Rohrlich}}, \
  and\ \bibinfo {author} {\bibfnamefont {E.~C.~G.}\ \bibnamefont {Sudarshan}},\
  }\href {\doibase 10.1063/1.1705135} {\bibfield  {journal} {\bibinfo
  {journal} {J. Math. Phys.}\ }\textbf {\bibinfo {volume} {8}},\ \bibinfo
  {pages} {2155} (\bibinfo {year} {1967})}\BibitemShut {NoStop}%
\end{thebibliography}%

\end{document}